\documentclass[journal,fleqn,10pt,twocolumn,twoside]{IEEEtran}

\usepackage{lipsum}
\usepackage{amsmath}
\usepackage{amssymb}
\usepackage{graphicx}
\usepackage{cite}
\usepackage{flushend} 
\usepackage{lastpage}
\usepackage{color}
\usepackage{xcolor,colortbl}
\usepackage{multirow}
\usepackage{caption}
\usepackage{subcaption}
\usepackage{comment}
 
\usepackage{tikz}
\definecolor{lime}{HTML}{A6CE39}
\DeclareRobustCommand{\orcidicon}{%
	\begin{tikzpicture}
	\draw[lime, fill=lime] (0,0) 
	circle [radius=0.16] 
	node[white] {{\fontfamily{qag}\selectfont \tiny ID}};
	\draw[white, fill=white] (-0.0625,0.095) 
	circle [radius=0.007];
	\end{tikzpicture}
	\hspace{-2mm}
}
\foreach \x in {A, ..., Z}{%
	\expandafter\xdef\csname orcid\x\endcsname{\noexpand\href{https://orcid.org/\csname orcidauthor\x\endcsname}{\noexpand\orcidicon}}
}

\usepackage{hyperref} 

\usepackage{tikz}
\usetikzlibrary{trees}
\usepackage[flushleft]{threeparttable} 
\usepackage[pscoord]{eso-pic}
\definecolor{ao}{rgb}{0.0, 0.0, 1.0}

\IEEEoverridecommandlockouts

\begin{document}

\title{A Survey on Applications of Cache-Aided NOMA}

\author{Dipen~Bepari\orcidA{},
Soumen~Mondal\orcidB{},
Aniruddha~Chandra\orcidC{},~\IEEEmembership{Senior~Member,~IEEE},\\
Rajeev~Shukla\orcidD{},~\IEEEmembership{Student~Member,~IEEE},
Yuanwei~Liu\orcidE{},~\IEEEmembership{Senior~Member,~IEEE},\\
\hspace{8mm}Mohsen~Guizani\orcidF{},~\IEEEmembership{Fellow,~IEEE}, and
Arumugam~Nallanathan\orcidG{},~\IEEEmembership{Fellow,~IEEE}%
\thanks{Manuscript received 29 April 2022; revised 31 October 2022 and XXX xx, 2023; accepted XXX xx, 2023. Date of publication XXX xx, 2023; date of current version XXX xx, 2023.}
%
%
\thanks{D. Bepari is with the Department of Electronics and Communication Engineering, National Institute of Technology Raipur, Chhattisgarh-492010, India (e-mail: dipen.jgec04@gmail.com).}
%
%
\thanks{S. Mondal, A. Chandra and R. Shukla are with the Department of Electronics and Communication Engineering, National Institute of Technology Durgapur, West Bengal-713209, India (e-mail: soumen.durgapur@gmail.com, aniruddha.chandra@ieee.org, rs.20ec1103@phd.nitdgp.ac.in).}
\thanks{Y. Liu and A. Nallanathan are with the School of Electronic Engineering and Computer Science, Queen Mary University of London, London, E1 4NS, U.K. (e-mail: yuanwei.liu@qmul.ac.uk, a.nallanathan@qmul.ac.uk).}
\thanks{M. Guizani is with the Machine Learning Department, Mohamed Bin Zayed University of Artificial Intelligence (MBZUAI), Abu Dhabi, UAE. (e-mail: mguizani@ieee.org). }
}

\maketitle
\markboth{IEEE COMMUNICATIONS SURVEYS \& TUTORIALS, Vol. XX, No. X, XXX QUARTER 2023}{BEPARI \MakeLowercase{\textit{et al.}}: A Survey on Applications of Cache-Aided NOMA}

\begin{abstract}
Contrary to orthogonal multiple-access (OMA), non-orthogonal multiple-access (NOMA) schemes can serve a pool of users without exploiting the scarce frequency or time domain resources. This is useful in meeting the future network requirements (5G and beyond systems), such as, low latency, massive connectivity, users’ fairness, and high spectral efficiency. On the other hand, content caching restricts duplicate data transmission by storing popular contents in advance at the network edge which reduces data traffic. In this survey, we focus on cache-aided NOMA-based wireless networks which can reap the benefits of both cache and NOMA; switching to NOMA from OMA enables cache-aided networks to push additional files to content servers in parallel and improve the cache hit probability. Beginning with fundamentals of the cache-aided NOMA technology, we summarize the performance goals of cache-aided NOMA systems, present the associated design challenges, and categorize the recent related literature based on their application verticals. Concomitant standardization activities and open research challenges are highlighted as well.
\end{abstract}

\begin{IEEEkeywords}
Caching, Non-orthogonal multiple access, Standardization.
\end{IEEEkeywords}

\section{Introduction} \label{sec:intro}
\IEEEPARstart{D}{espite} many challenges, there had been several successful trial-runs and limited-scale commercial deployments of the fifth generation ($5$G) cellular networks across the globe over the last couple of years \cite{Buchholz2021,Casetti2021}. $5$G implementations are, however, vastly heterogeneous as its three major use cases have conflicting requirements: enhanced mobile broadband (eMBB) promises Gbps connectivity on the go, massive machine type communication (mMTC) requires support for extremely high node density and low transmission power to enhance network lifetime, while ultra-reliable low-latency communication (URLLC) demands immediate response from a resilient network. Sixth generation ($6$G) networks aspire to touch all these three cornerstones, eMBB, mMTC and URLLC, all at once \cite{Dang2020}, but it is difficult to satisfy the data-rate, spectral-efficiency, and low-latency constraints, simultaneously. For example, a fully autonomous (level 5 autonomy) connected vehicle alone would generate $19$ terabytes (TB) of sensory data per hour \cite{Gotz2021}. If we consider the future internet-of-everything (IoE), including bio, nano and space domains, the data we need to transfer over the backhaul network in an hour can easily reach the order of zettabytes (ZB; $1$ ZB $\sim 10^9$ TB). The challenge is to deliver such a huge amount of data over limited-bandwidth links maintaining the strict latency constraints. $6$G aims to break the \emph{millisecond latency barrier} \cite{Panwar2020}, while some applications, such as haptic interactions through a tactile internet, demand an end-to-end delay even lesser than $1$ ms\cite{Xiang2019}. The pressure for a faster network is mounting as many game-changing device technologies are now mature enough to be prototyped: three-dimensional ($3$D) holographic displays are ready for thin gadgets \cite{Park2019}, virtual reality/ augmented reality (VR/AR) microscopes are detecting cancer cells \cite{Chen2019} and human body integrated wireless surfaces are being built for providing users with a truly-immersive extended reality (XR) experience \cite{Yu2019}. Development of these devices further intensifies the struggle for building an agile $6$G network. Undoubtedly, \emph{backhaul is our next frontier}. 

Cellular networks are more centralized than wired ones and \emph{caching} has been under consideration for improving the backhaul latency since the introduction of $3$G \cite{Erman2011}. Considering the fact that \emph{propagation delay is just one of the many components of the overall end-to-end delay}, and in addition there would be transmission delay (links are not of infinite capacity), buffering time (every in-between node has a finite storage capacity), multiple access delay (you are not the only one who is active), etc. the effective radius shrinks further down. Thus, either you have to bring the regional data centers (RDCs) to your locality or make sure the content is available (at least partially), in a local manner, i.e., perform caching.

The need for caching is more relevant than ever before with the paradigm shift in the internet protocol (IP) traffic pattern. In 2021, IP video consisted of $82\%$ of total internet traffic. The surge is due to the increasing popularity of all three types of video services, namely free (e.g., YouTube), subscription-based (e.g., Netflix) and social media (e.g., WhatsApp). In 2016, the video traffic consumed by mobile users was almost equal to the PC users. Five years later, the ratio is now heavily skewed, the mobile users are consuming videos $3$ times as much as the PC users. The request for specific high-quality multimedia content with low latency, irrespective of users’ locations, converted the communication-centric networks to content-centric networks. A major amount of backhaul traffic is due to frequently transmitting replica of the same content (say, Despacito by Luis Fonsi or Baby Shark Dance). Roughly $5\%$ of the webpages, audio and video files are popular, and a large number of users request these popular files at different time instants, impelling the network to provide the same content, again and again, using the backhaul link. 

\begin{table*}[h t b! ]
  \renewcommand{\arraystretch}{1.3}
    \caption{\small Timeline of existing surveys on NOMA and caching. }\label{Table: exist_survey}
  \centering
\begin{tabular}{|p{08 mm}| p{5 mm} |p{6 mm}| p{54 mm }| p{35 mm }| p{49 mm}|} 
  \hline \hline
    \rowcolor{lightgray}
    \centering \textbf{Tech.} &\textbf{Year} &\centering \textbf{Ref.}& \centering \textbf{Focus area} & \centering \textbf{Metric} &\qquad \textbf{Application} \\
\hline
\hline 
\multirow{9}{*}{NOMA} & 2016 & \cite{Wei2016ASurvey} &  Single carrier vs. multi-carrier NOMA  & Sum rate & MIMO, CoCom\\ 
\cline{2-6}
 & 2017 & \cite{ Ding2017Asurvey} &  Single carrier vs. multi-carrier NOMA & Channel gain, Sum rate & MIMO, CoCom, mmWave \\
 \cline{2-6}
 & 2018  & \cite{ Dai2018 } &  PD-NOMA vs. CD-NOMA & Spectral efficiency, Receiver complexity & LTE, 5G\\
  \cline{2-6}
  & 2018  & \cite{ aldababsa2018tutorial } &   NOMA vs. OMA & Sum rate, Outage probability & MIMO, CoCom \\
  \cline{2-6}
   & 2019  & \cite{ Vaezi2019Interplay } &  Coexistence of NOMA with other technologies & UL and DL architecture & MIMO, CoCom, mmWave, CR, VLC \\
  \cline{2-6}
  & 2020 & \cite{ Omar2020A} &   PD-NOMA integration with other technologies  & Optimal rate & MIMO, mmWave, CoMP, CR, VLC, UAV{} \\
  \cline{2-6}
  &2020 & \cite{ Makki2020ASurvey } &  Adoption of NOMA in future standards & Energy efficiency, end-to-end delay & 5GNR \\
  \cline{2-6}
  & 2021 & \cite{ Akbar2021A } &  Security and resource allocation & PHY security, User pairing, Power allocation & MIMO, CoCom, CR, UAV, SWIPT \\
  \cline{2-6}
  & 2022 &\cite{ yahya2022error } &  Error rate calculation & BER, SER, PEP, BLER & CoCom, VLC, FSO, IRS \\
  \hline   
 \multicolumn{3}{|c|}{ \begin{tabular}[t]{@{}c@{}}Cache-aided NOMA\\(This paper)\end{tabular}}  &   Interplay of cache and NOMA & Sum rate, Delay, Decoding probability & MIMO, mmWave, SWIPT, IRS, V2X\\
\hline
\hline 
    \rowcolor{lightgray}
    \centering \textbf{Tech.} &\textbf{Year} &\centering \textbf{Ref.}& \centering \textbf{Focus area} & \centering \textbf{Architecture} &\qquad \textbf{Application} \\
\hline
\hline   
\multirow{8}{*}{Cache} & 2012 & \cite{passarella2012survey} & Caching in P2P  & Web caching & CDN, P2P\\
\cline{2-6}
  & 2013 & \cite{xylomenos2013survey} &  Caching in international ICN projects & On-path, Off-path & ICN \\
  \cline{2-6}
  & 2013 & \cite{zhang2013caching} &  ICN caching vs. web/P2P caching & ICN caching & ICN \\
  \cline{2-6}
   & 2018 &\cite{Wang2018Integration} &  Integration of networking, caching, computing  & D2D, Cooperative &  Wireless, Cloud, MEC \\
  \cline{2-6}
  & 2018 & \cite{li2018survey} &  Cache enabled cellular networks   & SBS, MBS, D2D & 5G \\
  \cline{2-6}
  & 2020 &\cite{Zahed2020A}&  Energy efficient caching  & SBS, MBS, D2D & ICN \\
  \cline{2-6}
  & 2020 &\cite{kabir2020role} & Cache utility to network operators & SBS, MBS, D2D  & 5G \\
  \cline{2-6}
  & 2022 &\cite{al2022caching }&  Caching in IoT  & ICN caching & IoT   \\
\hline  
\multicolumn{3}{|c|}{ \begin{tabular}[t]{@{}c@{}}Cache-aided NOMA\\(This paper)\end{tabular}}  &   Interplay of cache and NOMA & D2D, Cooperative, Edge & MEC, CFmMIMO\\ 
 \hline
 \hline
\end{tabular} 
\end{table*}

 Caching can reduce both backhaul use and latency; popular contents, asked by the users frequently, are stored near the network edge (e.g. at base stations, users’ device) in advance during the off-peak period. When users request a common file, the network delivers the file from the cache without engaging the backhaul infrastructure all the way back to core. To store the popular contents in the cache, the network needs to access the backhaul links only once, thus avoiding accessing backhaul multiple times during peak hour \cite{ Zeydan-16}. Unlike, bandwidth and power, which are limited communication resources, content caching resources are adequately available, cost-effective, and suitably maintainable. Moreover, caching resources are growing following the Moore’s law. \emph{Installing memory for caching is cheaper than that of increasing backhaul capacity}; the retail price of a 2-3 TB memory is approximately 100 USD \cite{Golrezaei13}. The non-causality characteristic of caching operation is particularly useful for mobile networks; in highly mobile 5G wireless environments caching at user equipment (UE) not only enhances the video streaming quality but also reduces the number of handovers, mitigates handover failure, and decreases energy consumption \cite{Semiari2018Caching}.
 
The cache technique is greatly compatible with many advanced communication systems, like millimeter-wave (mmWave) communications \cite{Semiari18Caching, Hao20Edge}, multiple-input multiple-output (MIMO) systems \cite{Cao17Fundamental, Garg21Function}, Mobile edge computing (MEC) \cite{Wang19In, Tran19Adaptive}, terahertz communication \cite{zhang2020energy} and others. However, our survey focuses on the cache-aided non-orthogonal multiple access (NOMA) technique. NOMA is one of the promising technologies for next-generation wireless communication \cite {Liu-17, tse2005fundamentals}. It is capable of efficiently realizing higher system throughput and spectral efficiency compared to the traditional orthogonal multiple access (OMA) \cite {Saito2015}. Maintaining the fairness of users, NOMA serves multiple users simultaneously at the same frequency band/ time/code. The key idea of NOMA is to apply superposition coding at the transmitter for combing the signals of multiple users and successive interference cancellation (SIC) method at the receiver for decoding individual signal \cite{Dai2018, Ding2017Asurvey, Liu2017}. The SIC operation is a complex procedure, specifically, when AP serves a large number of users within a time-frequency block \cite{ yue2018unified }. Cache-aided NOMA receivers exploiting the cache, when requested files of the other users are fully/partially available, can apply the cache-enabled interference cancellation (CIC) process and reduce the complexity of the SIC process \cite{Yang2017Joint}. A fundamental concept of the NOMA technique and its application in long-term evolution (LTE) and 5G have been reported in \cite{Saito2013, Ding17}. It is also reported that the amalgamation of NOMA with cache technology can achieve a significant system performance enhancement. A cache-aided NOMA network reaps benefits from both the cache and NOMA techniques.

\subsection{Motivation} \label{sec: motivation}
Some of the existing literature explore various insights of cache strategies and NOMA principles separately. Table \ref{Table: exist_survey} illustrates the primary focus of the survey papers on NOMA and caching techniques. These  articles analyze different features of NOMA and cache individually. Along with the state-of-the-art NOMA techniques for future networks (e.g., 5G and beyond system), the fundamental operating principles of NOMA and their comparative performance analysis over the OMA are the primary focus of \cite{ Dai2018 }. Ding \textit{at el.} provided a broader overview of research innovation of NOMA and their applications in advanced communications along with associated implementation challenges and constraints \cite{ Ding2017Asurvey, Akbar2021A }. The resource allocation and the performance analysis of MIMO-NOMA systems and the comparison between  Welch-bound equality spread multiple access (WSMA)-based NOMA and multi-user-MIMO are reported in \cite{aldababsa2018tutorial} and \cite{ Makki2020ASurvey}, respectively. Vaezi \textit{at el.} \cite{ Vaezi2019Interplay} analyzed an interplay between NOMA and other technologies, like MIMO, massive MIMO, mmWave communications, cognitive communications, visible light communications, etc., and summarized how these combined technologies elevate network performance in terms of scalability, spectral efficiency, energy efficiency, etc.. The survey paper \cite{ Omar2020A } analyzed various optimization scenarios to investigate only the maximum achievable sum-rate when PD-NOMA amalgamates with other promising technologies for 5G and beyond 5G (B5G) and ignored the analysis of other crucial performance metrics. The authors in \cite{Vaezi2019Interplay, Omar2020A}, completely overlooked a systematic analysis of wireless networks combining NOMA and cache technologies. The authors explored a comparative study of various adopted approaches (such as optimization techniques, analytical methods, game theory, matching theory, graph theory, machine learning techniques, etc.) to address the problem of resource allocation, signaling, and practical implementation of NOMA technologies in \cite{ Akbar2021A }. In this paper, the authors analyze the security aspects of NOMA technologies. Recently, Yahya \textit{at el.} first enlightens the error rate performance of various NOMA configurations and designs in a holistic manner \cite{ yahya2022error }. 
 
In \cite{Zahed2020A}, the authors studied recent development on the green content caching technique to explore various cache-equipped wireless networks, research-gap, solution methods, and application areas. In \cite{ li2018survey}, Liying \textit{et al.} presented a fundamental concept of caching techniques and their recent development in various types of cellular networks such as macro-cellular, heterogeneous, device-to-device, cloud-radio access, and fog-radio access networks. A few articles also studied the caching technique in cellular systems \cite{Ji16Wireless, Wang2018Integration }. In \cite{Wang2018Integration}, the authors present the research challenges of cache-aided integrated networking in wireless communication systems. The survey \cite{ Steve2017ASurvey } focused on caching techniques for vehicular communications. Although individual surveys on NOMA and caching do exist \cite{Dai2018, Zahed2020A, li2018survey}, explicit analysis of cache-aided NOMA systems and their applications have not been reported yet.

\subsection{Contributions}
The primary goal of this survey is to present a systematic study of the recent research development and innovations in the cache-aided NOMA systems. Numerous research articles analyze cache and NOMA-based wireless networks individually and exploit their benefits. However, to the best of our knowledge, this is the first article that introduces a survey on caching-aided NOMA systems and their practical applications in 5G and beyond systems. After a brief tutorial on the concept of wireless caching and NOMA, we explained the integration of NOMA with wireless caching by elaborating the underlining design principles, features and key performance indicators. We also presented a formal classification of cache-aided NOMA systems based on their diverse practical applications through highlighting the state-of-the-art, associated challenges, and promises. The forthcoming networks require massive data traffic to be carried over backhaul links. In this regard, we explored the fundamental impacts of cache-aided NOMA systems in terms of network efficiency, QoS and latency. Furthermore, this article presents a detailed account of the standardization activity and real-time development news of NOMA and cache technologies. Finally, this article identifies a wide range of potential future research opportunities and related technical challenges that need to be addressed for implementing cache-aided NOMA systems.

\begin{figure*}[t]
\begin{center}
\includegraphics[ scale=0.30]{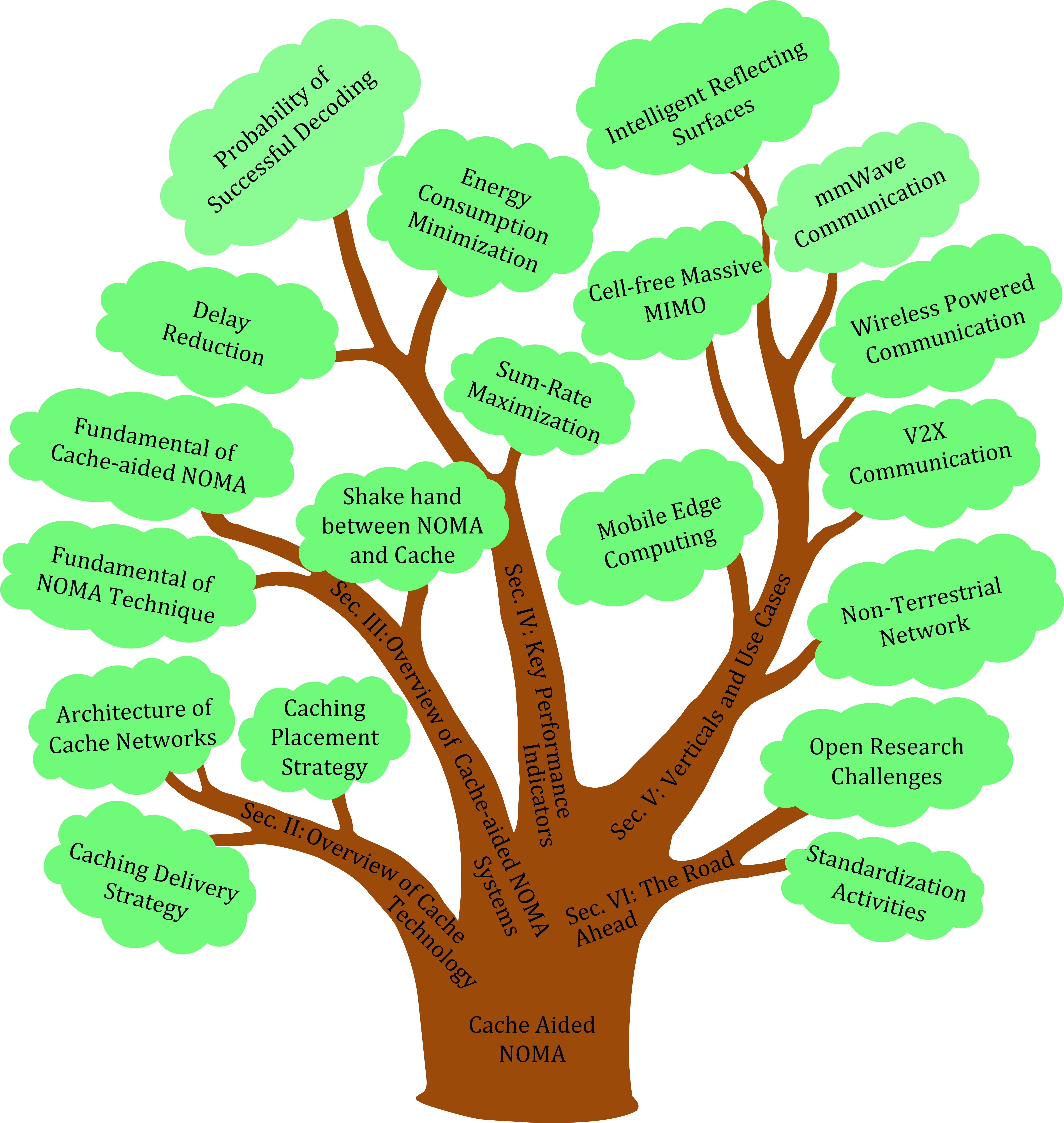}
\caption{Organization of this survey.}
\label{Fig: taxo}
\end{center}
\end{figure*}

\subsection{Organization}
The organization of this paper is shown in Fig. \ref{Fig: taxo}. Section \ref{sec: Overview_Cache} provides an overview of cache technology. Section \ref{sec: Overview_CacheNOMA} and Section \ref{sec: KPI} present an overview of cache-aided NOMA systems and explores various key performance indicator of cache-aided NOMA systems respectively. Next, Section \ref{sec: application} focuses on the application of cache-aided NOMA in the wireless communication domain, i.e., non-terrestrial networks, MEC, vehicle-to-everything (V2X) communication, cell-free massive MIMO, mmWave communications, etc. Section \ref{sec: open} highlights standardization activities of cache-aided NOMA and identifies possible directions for future research on cache-aided NOMA. Finally, the conclusion of the paper is presented in Section \ref{sec: conc}. A list of acronyms is tabulated in Table \ref{Table: acronyms}. 

\begin{table}[h t b!]
\renewcommand{\arraystretch}{1.3}
 \caption{\small List of acronyms frequently used in this survey}
  \label{Table: acronyms}
  \centering
 \begin{tabular}{|p{15 mm} | p{62mm} | }
  \hline
  \rowcolor{lightgray}
\textbf{Acronyms}  &  \textbf {Description}   \\ 
\hline
\small
AP & Access point\\
\hline
B5G & Beyond Fifth-Generation \\
\hline
CFmMIMO & Cell-free Massive Multi-Input Multiple-Output\\
\hline
CIC & Cache-enabled interference cancellation\\
\hline
CoCom & Cooperative Communication\\
\hline
CSI & Channel State Information\\
\hline
D2D  & Device-to-Device \\
\hline
ETC & Edge Caching Technique  \\
\hline
eMBB & Enhanced Mobile Broadband\\
\hline
ICN &  Information Centric Networking  \\
\hline
IoT & Internet of Thing\\
\hline
LFU & Least Frequently Used \\
\hline
LRU & Least Recently Used \\
\hline
LTE & long Term Evolution \\
\hline
MBS & Macro Base Station\\
\hline
MEC & Mobile Edge Computing\\
\hline
MIMO & Multiple-Input Multiple-Output \\
\hline
mMTC & Massive Machine Type Communication\\
\hline
mmWave & Millimeter-Wave\\
\hline
MUSA & Multiuser Shared Access \\
\hline
NOMA & Non-Orthogonal Multiple Access\\
\hline
OMA & Orthogonal Multiple Access\\
\hline
OTFS & Orthogonal Time-Frequency Space\\ 
\hline
 PD-NOMA & Power-Domain NOMA\\
\hline
PER & Poll-Each-Read \\
\hline
QoS & Quality of Service \\
\hline
SBS & Small-cell Base Station\\
\hline
SCMA & Sparse Code Multiple Access\\
\hline
SDP & Successful Decoding Probability\\
\hline
SIC & Successive Interference Cancellation\\
\hline
SWPT & Simultaneous wireless information and power transfer\\
\hline
UAV & Unmanned Aerial Vehicle \\
\hline
UE &  User Equipment\\
\hline
URLLC &  Ultra-Reliable Low-Latency Communication\\
\hline
V2N & Vehicle-to-Network \\
\hline
V2V & Vehicle-to-Vehicle \\
\hline
V2X & Vehicle-to-everything\\
\hline
VLC & Visible Light Communications \\
\hline
\end{tabular}
 \end{table}

\section{Overview of Cache Technology}\label{sec: Overview_Cache}
The types of demand for content are changing day by day. In the early 1990s, Web pages and images were in high demand, and excess delivery of these contents was responsible for heavy network congestion. To cope with this problem Web caching technique becomes a promising approach for significantly reducing traffic load \cite{Wang-1999, ali2011survey }. In the early 2000s, the primary reason for network congestion was due to the demand for video content, and that was challenged by caching for content distribution networking \cite{passarella2012survey} and information-centric networking (ICN) \cite{ zhang2013caching}. In ICN, the distance between cache and user was further shortened, and a new feature, \textit{popularity of contents}, was incorporated in the content placement techniques \cite{ xylomenos2013survey}. Nowadays, users created video files and their delivery introduces additional network traffic load over wireless channels. Establishing a wireless caching network is more challenging than wired caching networks because of the unpredictable movement of the users and uncertain channel gain quality. Since 2009, researchers showing their interest in wireless caching for reducing traffic load and increasing the quality of communications \cite{ Niesen-2009, Keng-Tec-2009}. In 2010, the authors confirmed that the caching technique enhances the system throughput by as much as 400--500\% \cite{WebCache}. The focus of the research was primarily limited to the development of the cache infrastructure, gateway design, routing, cooperative, and physical layers. However, after 2010, the researchers combined cache technology with other technologies and analyzed their performances in various wireless networks for next-generation communications.\\

 The latency minimization is one of the stringent requirements and significant challenges for next-generation intelligent applications. Let us review the importance of caching from the $5$G latency requirement perspective, where the projected latency is not more than $0.5$ ms \cite { multiple2016}. Electromagnetic (EM) waves travel at $3 \times 10^5$ km/s ($=c$) in an unguided medium and can cover a round-trip distance of $150$ km over free space in $1$ ms. However, most backhauls are not wireless, they are almost always created with the optical fiber. For a single-mode fiber having a core refractive index of $1.49$ $(=n)$, the velocity of light propagation becomes $v=c/n=0.67 c$, reducing the coverage to $100$ km. Unless we invent a carrier that travels faster than light, regional data centers (RDCs) cannot be further away. Caching can be a great approach to minimizing the latency, where a few most popular contents are stored in a cache memory before being asked by users. The major challenges of wireless caching are:
\begin{itemize}
\item A wireless channel bandwidth is very limited compared to wired channels.
\item A wireless channel gain quality is very poor due to noise, interference, shadowing, and so on. 
\item Wireless devices may be disconnected because of high mobility or/and poor channel gains. 
\item Limited battery life restricts the increase of transmit power.  
\item Limited cache memory demands efficient cache placement and access strategies. 
\end{itemize}

The caching strategy involves two operating phases, the content placement phase, and the content delivery phase. In the content placement phase, the network stores popular contents in the cache memory during off-peak time. In the content delivery phase, the network serves the cached contents during peak traffic hours. Efficient caching placement with an updated strategy and delivery strategy is required to get maximum benefits from the caching strategies.

\subsection{Caching Placement Strategy}
Appropriate content placement is the baseline for achieving a significant performance gain from any cache-aided system. The caching placement strategy determines the size and location of files and decides how and where the selected files are to be downloaded to the cache memory. A content replacement mechanism that determines how to update the cached content regularly is an integral part of the cache placement strategy. Though a network requires a minimal up-gradation in the existing infrastructure for caching, significant challenges are involved with caching strategies. Because of variable content sizes, random demand for contents, limited cache resources, and the movement of the users, cache management becomes a challenging issue. Accommodating large numbers of files in the cache with limited memory space is one of the most severe issues.

The popularity of the files has to consider in the cache placement strategy for effectively reducing the use of the backhaul link in cache-aided systems. Increasing the availability of the requested files as much as possible is a key factor. Unpopular content selection for caching may lead to a considerable overhead cost \cite{Nuggehalli2006Efficient}. The popularity of the randomly requested contents is widely modeled by the Zipf distribution  \cite{ Zipf1, Zipf2, Zipf3}. The widely used Zipf distribution is proficient for measuring the polarity of video files \cite{ Zipf2 }. In \cite{ Hachem2017Coded }, based on the varying degree of popularity, the authors proposed a multiple-level non-uniform content popularity in their research work.

Two types of content placement strategy broadly found in literature- \textit{coded placement strategy} \cite{CodedCache1, CodedCache3, CodedCache4, TradeoffCache, Jesper19} and\textit{ uncoded placement strategy} \cite{Rao2016Optimal, dimokas2008cooperative, Xiuhua2016Weighted}. The basic principle of coded placement strategy is to divide the files into multiple small segments, encode the segments by a coding methods and place them in the cache memory. During the content delivery phase, a certain coding technique needs to employ to combine the requested files. Raptor codes \cite{RaptorCodes} and fountain codes \cite{FountainCodes} are popularly used for combining the file segments. The uncoded placement strategy is comparatively simple where complete requested file or a portion of the file is kept in the cache.   

The coded cache-enabled system with $K$ users, cache capacity of $F$ files, and $N$ available files in the cache achieve a caching gain of $\frac{1}{1+\frac {KF}{N}}$ over the un-cached system \cite{ CodedCache3}. The gain indicates a large amount of rate reduction in the shared link. The coded placement strategy is suitable for reducing cache memory consumption with increasing computational complexity, specifically for a large number of segment files \cite{CodedCache1}. The authors in \cite{ altman2013coding }, explore the advantages of coded caching strategies when cache-enabled access points (APs) like BS, SBS, MBS, etc.) are randomly distributed. In \cite{Ostovari2013Cache}, authors have proposed a linear network coding-based cache content placement strategy that increases the amount of available data compared to triangular network coding. The coded caching strategies exploit coded multicast opportunities that further decrease the backhaul traffic, particularly for densely deployed cache-enabled APs \cite{ Zhang2018Coded }. Based on the random caching placement and multiple group cast index coding, the authors have proposed an order-optimal coded caching placement \cite{ Ji2015Caching }. Niesen \textit{et al.} \cite{ Niesen2017Coded} verified that optimal cache placement minimizes the traffic load of the shared link knowing the popularity distribution of files. Binbin \textit{et al.} implemented an optimal cache content placement in cache-aided BS to reduce the backhaul traffic load of a wireless access network \cite{ Dai2016Joint }. Jinbei \textit{et al.} \cite{ Zhang2018Coded } presented an analysis that finds the lower bound on the data transmission rate of any coded caching strategy. Furthermore, for any popularity distributions and the system size, the authors also derived a constant factor that indicated the gap of achievable average transmission rate from the optimal.

A cache replacement strategy is accompanied by the cache placement strategy. It is an essential mechanism needed to employ for cache-aided systems. The least recently used (LRU) and least frequently used (LFU) are the traditional replacement policies found in the literature \cite{ Podlipnig-2003, Balamash-2004, Wang-1999, Yi-Bin-2003}. The LRU replaces the least recently used files, and the LFU replaces the least frequently used files. The combination of cache access and replacement strategy makes a caching method. Various cache access and placement strategies and their performances are analyzed in \cite{ Chen-06 }. A gain-based cache replacement policy named, Min-SAUD \cite{ Xu04 } and a hotspot-based caching scheme \cite{ Zhang-2019 } are adopted to satisfy the caching replacement requirements. In \cite{ lei2019optimal}, authors have proposed a cache replacement and content delivery strategy for regularly updating cached contents.  

\subsection{Caching Delivery Strategy}
The availability of the requested files in the associated cache is not the only event that enhances the system performance, users need to receive and decode requested files in an error-free manner. In addition, a caching delivery strategy determines where the requested file will be transmitted from, the transmitting frequency band, the transmission power, and the encoding process so that files arrive at the requested user successfully and quickly. Poll-Each-Read (PER), Call-Back (CB), and Invalidated Report (IR) are some of the classical cache access schemes. BSs need to employ coding methods for coded transmission to combine user-requested files, whereas BSs deliver cached files individually for uncoded transmission techniques. Whenever any user requires a file, the network first searches it in the cache memory before downloading it from the internet server. Searching files in the cache may take substantial time, especially when the \textit{miss rate} (cache memory fails to provide a shout file) is high due to a shortage of cache memory or/and storing less popular content in the cache \cite{AAT}. Therefore, overall system performance primarily depends on how efficiently popular files are cached.

The backhaul traffic load and content delivery latency of cache-aided networks are subject to the cache memory size. The fundamental trade-off between the advantage of caching and the cache storage capacity has been studied in \cite{ TradeoffCache, Tradeoffs } for coded and uncoded cache systems. They have analyzed the trade-off from an information-theoretic perspective, and carried out investigation based on the \textit{normalized delivery time} metric, which measures the worst-case content delivery time subjected to the transmission rate of the requested files. Aiming to maximize the successful download probability of requested files, authors in \cite{ Peng16 } have optimized the cache memory size subjected to channel statistics, backhaul capacity, and distribution of file popularity in a  cellular network. In \cite{xiang2019cache}, the authors have formulated an NP-hard optimization problem to minimize the time required to complete a file delivery for downlink transmission of a cache-aided network. Average latency in both the backhaul and cache link for delivering the requested file of a cache-enabled system under the constraint of quality of the recommended files has been minimized in \cite{ fu2021efficient}. Various types of caching strategies have been proposed, and the performance of these methods is analyzed mostly in terms of a cache hit rate.

\subsection{Architecture of Cache Networks}
 Depending on the infrastructure for the downlink data transmission mechanism, the wireless caching network architectures are grouped into two categories. The first category is device-to-device (D2D) caching and the second category is edge caching.

\subsubsection{D2D Caching Technique}
\begin{figure} [t]
	\begin{center}
		\includegraphics[scale=0.70]{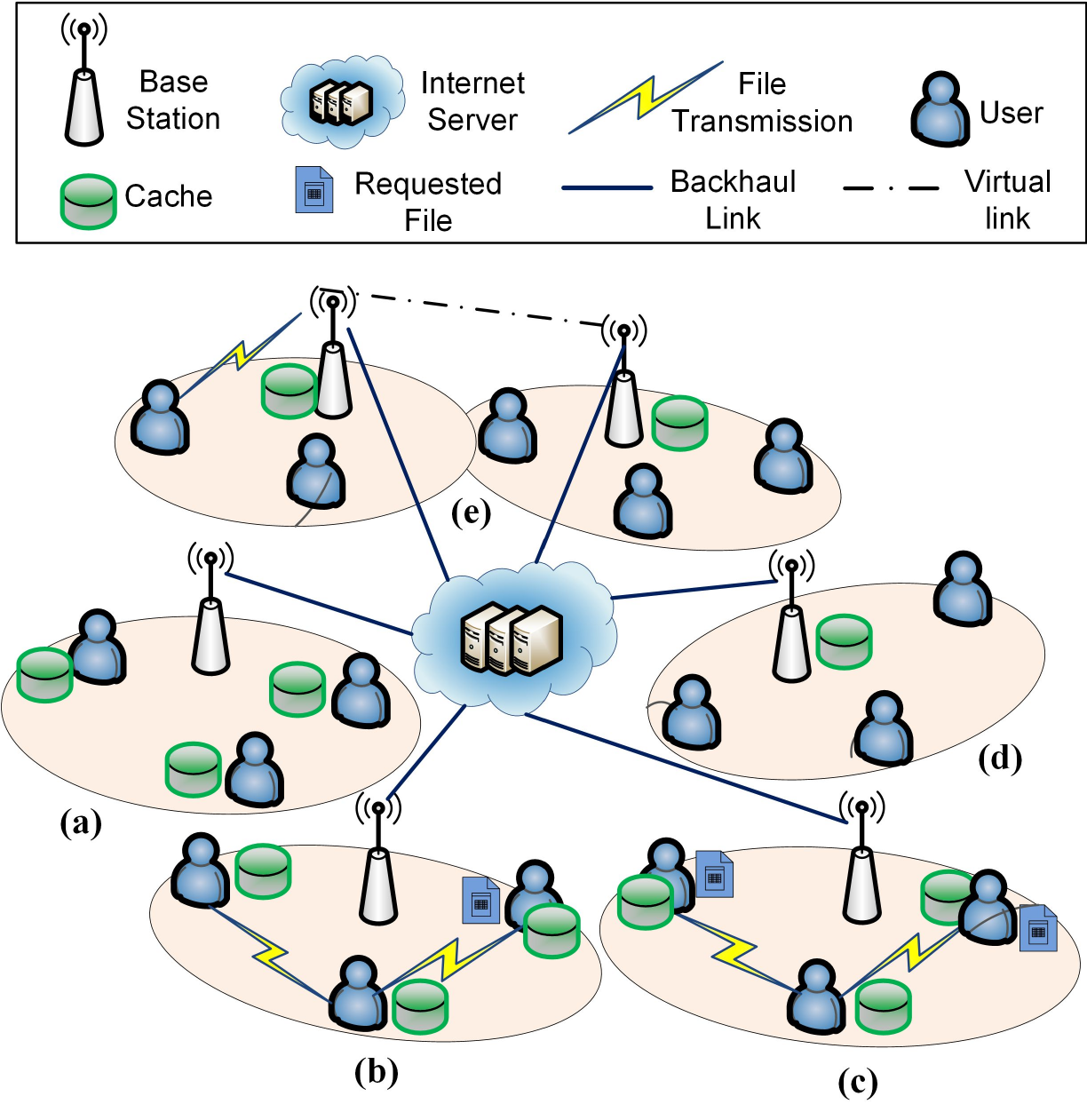}
		\caption{\small Various architectures of cache networks (a) D2D caching technique, (b) D2D multihop relay, (c) Cooperative D2D, (d) edge caching technique, and (e) Cooperative edge caching technique}
		\label{Fig.NetworkArchitech}
	\end{center}
\end{figure}

In D2D caching, shown in Fig. \ref{Fig.NetworkArchitech}(a), dedicated infrastructure for caching is not available. Users depend on the content stored by neighbors \cite{Cheng17, D2Dsurvey,  Wang2017Mobility}. During the content placement phase, users store a few contents in their device, and during the content delivery phase, users communicate with the internet server through BS only when none of its neighbors has cached the requested file. A high density of users increases the availability of requested content at the nearby cache devices. D2D primarily relies on the cooperation of the neighboring users. D2D caching mainly improves spectral efficiency. Two types of D2D caching networks are found in the literature,\textit{ D2D Multihop relay} and \textit{Cooperative D2D}.\\ 
\textit{D2D Multihop relay:-} In D2D communication, intermediate users help to deliver the file from cache to the destination, as shown in Fig. \ref{Fig.NetworkArchitech}(b). Multihop delivery empowers a user to access the desired file from a cache-user located far away \cite{Ji13Optimal,kim2009active}.\\
\textit{Cooperative D2D:-} {When a user requested file is available to multiple nearby cache the file can be transmitted cooperatively to the destination,} as shown in Fig. \ref{Fig.NetworkArchitech}(c). In this case, the implementation of MIMO technology accelerates the file transmission process. In \cite{taghizadeh2010cooperative}, authors have considered a cooperative D2D caching for a wireless sensor network.

\subsubsection{Edge Caching Technique}
Unlike D2D Caching, in edge caching, dedicated caching infrastructure is available at the AP \cite{ma2020enabling, Yang2020Cache, EdgeCaching }, as shown in Fig \ref{Fig.NetworkArchitech}(d). In the content placement phase, popular contents are timely stored in the cache memory with high reliability before users ask for it. The primary aim of the content delivery phase is to provide the requested files without communicating with the internet server, thus enabling improvement in delivery latency. The latency of this technique is lower than that of in multihop transmission method. The edge caching technique primarily reduces the backhaul cost.

\textit{Cooperative Edge Caching Technique:-} In this case, BSs communicate with the neighboring BS for the requested contents. When a user requested file is available neither in nearby cache device nor in its BS, the request is forwarded to the neighboring BS, and on the availability of the file in cache of the neighboring BS, the user can get its file via its own BS. Figure \ref{Fig.NetworkArchitech}(e) shows the cooperative BSs delivery. Shan \textit{et al.}, \cite{shan2021performance} analyses the performance of a cooperative edge caching in wireless cellular networks. 
 
In the D2D networks, the users' devices are equipped with a cache, whereas in edge caching, APs are equipped with a dedicated cache facility. A few of the literature have considered hydride wireless caching networks, where cache infrastructure is available at both the transmitter (BS, SBS, and access points) and receiver (user) end \cite{ TradeoffCache, Hachem18}. During the content placement phase, files are stored individually in the transmitter and users' devices, and during the content delivery, the BS searches its own and users’ caches for the asked file before downloading from the internet servers. Fundamentally, edge caching is a centralized caching technique where based on the content popularity and network parameters, a BS decides which files at what time and where to be cached, and the BS also decides the requested file delivery strategies \cite{Zipf1}. On the other hand, D2D caching is a distributed caching technique where each device determines cache placement and delivery strategy. Distributed caching algorithms are comparatively lesser complex than centralized ones but may fail to provide global optimality of the algorithms.

\textbf{Summary:-} The first and foremost task of establishing cache-aided networks is to make available the most popular contents in the cache before being requested. The Zipf distribution is the most commonly adopted approach for modeling the popularity of randomly requested contents. The popularity of the content, distribution of content popularity, correlation among contents, location of users, and finite storage availability constraints are required to be taken into consideration in the caching strategy, which introduces challenges to the caching strategies. Furthermore, a wide range of variety in the content and sudden change in popularity intensify difficulties and show the drawback of content placement during \textit{off-hours}. Two online popularity prediction strategies, named the popularity prediction model (PPM) and the Grassmannian prediction model (GPM), were proposed to predict popularity in advance \cite{Garg2020Online}. The rise of the content day by day may change statistical values of popularity distributions and content popularity. Therefore, caching demands an efficient cache placement strategy to encounter these changes, which increases challenges in designing such an algorithm \cite{ Wang2014Cachein }. Although codded caching techniques demand additional coding overhead, but extraordinarily enhance system performance in terms of reducing bandwidth requirements and transmission latency compared to that of the best-uncoded technique \cite{CodedCache1}. The researchers assume error-free channels during content pushing, which is questionable in real-time communication scenarios. Therefore, a more practical channel model needs to be taken into account during content placement. 


\section{Overview of cache-aided NOMA Systems}\label{sec: Overview_CacheNOMA}
The primary aim of this section is to provide a brief overview of the NOMA operation and challenges associated with the cache-aided NOMA.     

\subsection{Fundamental of the NOMA Technique} 
With the increase of wireless devices, the multiple access (MA) technique becomes a popular approach to meet the demand of large numbers of wireless users. The MA techniques accommodate multiple users within the same resource blocks like frequency band, time slot, or spatial direction, concurrently. The MA can be categorized as OMA and NOMA. In OMA, resources like time, frequency, and code are orthogonal which reduces the interference introduced by other users. With the further advancement of wireless communication to fulfill the demand of massive connectivity, NOMA allows multiple users to use non-orthogonal resources simultaneously \cite{ding2017application}. 

\begin{figure}[t]
\begin{center}
\includegraphics[scale=0.54]{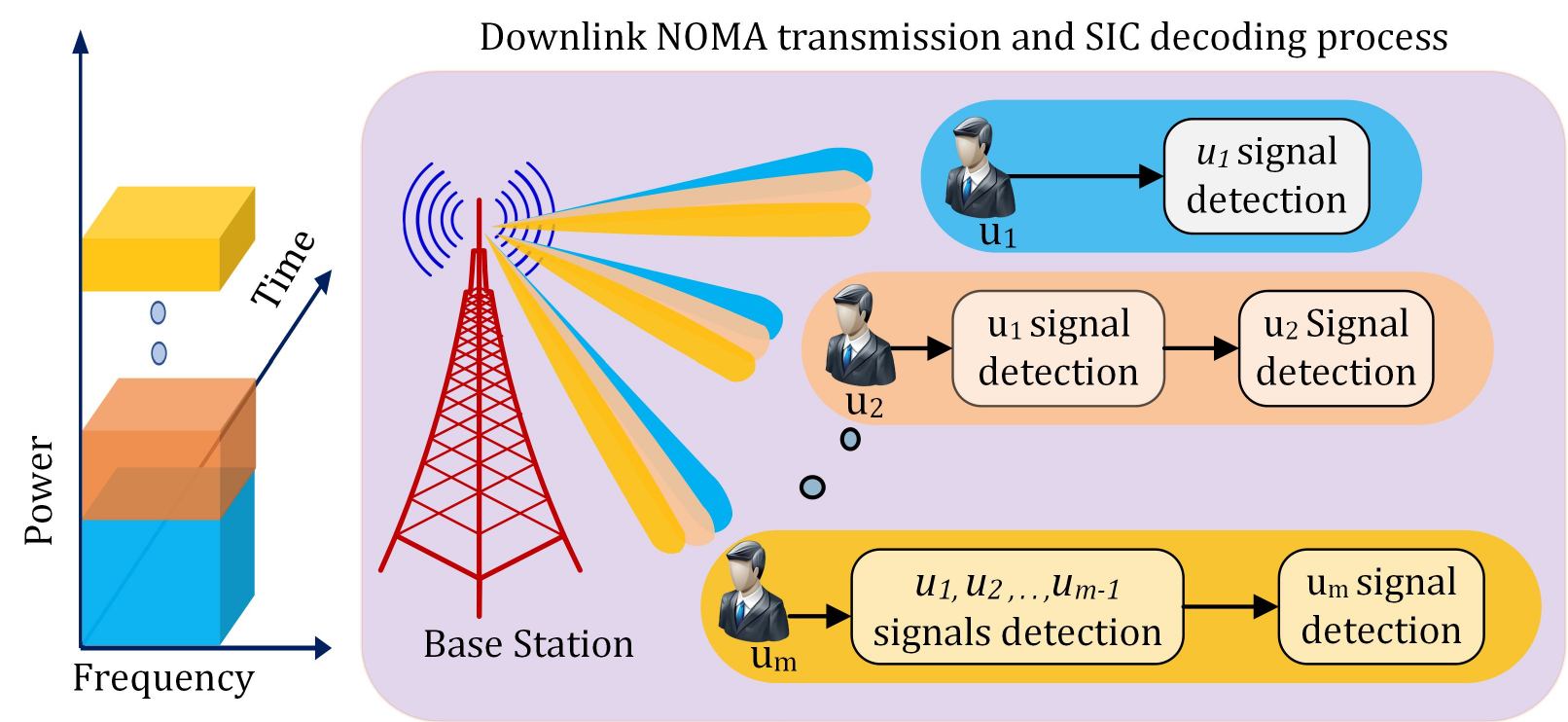}
    \caption{\small Power distribution strategy for PD-NOMA. The transmitter allocates maximum power to user $u_1$ with the weakest channel and minimum power to user $u_m$ with the strongest channel. $u_1$ decodes its own signal directly, and $u_m$ first decodes signal of $u_1, u_2,...,u_{m-1}$ then its own signal.}
 \label{Fig: noma_basic}
\end{center}
\end{figure}

The power-domain NOMA (PD-NOMA) and code-domain NOMA (CD-NOMA) are two main types of NOMA. In PD-NOMA, a transmitter transmits signals to multiple users exploiting the power domain. The transmitter combines signals of multiple users with different power levels by applying superposition coding. Unlike the water-filling algorithm for power allocation in OMA \cite{Bepari2015ImprovedPL, bepari2016spectral}, in PD-NOMA, the total power is distributed among the users in such a way that signals of users with weaker channel conditions comparatively get more power than the signals of users with stronger channels. As a consequence, (i) a stronger user achieves a higher data transmission rate with low transmit power, and (ii) a weaker user experiences limited interference caused by stronger users, and simultaneously, higher power allocation to weaker users improves users' fairness, spectral efficiency, and sum-rate \cite{men2015performance, ding2015impact}. At the receiver end, the receiver applies the SIC process to decode its signal. In the SIC process, the receiver first decodes the signal which has the highest assigned power and then subtracts the signal from the received composite signal. This process continues until the signal of the intended user is decoded \cite{yang2017optimality}. The power allocation and decoding process of PD-NOMA is shown in Fig. \ref{Fig: noma_basic}.  

The CD-NOMA assigns different codes to users and superimposes them over the same time and frequency. The operations of CD-NOMA are quite different from the PD-NOMA. The CD-NOMA is fundamentally built up on the concept of allocating different spreading codes that enable signals separation at the receiver. The multiuser shared access (MUSA) \cite{yuan2016non}, sparse code multiple access (SCMA) \cite{nikopour2013sparse}, low density spreading (LDS) \cite{hoshyar2008novel} are the main three types of CD-NOMA. In CD-NOMA, each user is supplied a codebook consisting of codewords. The transmitter first encodes the data of the user-requested content, then maps them into a complex codeword selected from the codebook and transmits over the channel allocating equal power to every user. The codebook of each user acts as a signature of the corresponding user. The BS delivers the user-requested content using CD-NOMA from the cache when it is available. Otherwise, the BS fetches it from the content server and then transmits it using the CD-NOMA principle. Though the CD-NOMA can remarkably enhance spectral efficiency, it requires a wide transmission bandwidth and considerable modification to the existing communication systems. On the contrary, PD-NOMA neither requires a major upgrade to the present communication networks nor a high transmission bandwidth \cite{ Dai2015Non}. In addition, PD-NOMA has a low complexity system compared to the CD-NOMA from a design perspective.  
 
It is necessary to know the order of the average channel gains or instantaneous channel state information (CSI) for fixed power allocation \cite{liu2018decode} or dynamic channel allocation \cite{yang2017impact}, respectively. The users' ordering is accomplished mostly  based on the instantaneous value of the CSI \cite {mondal2020outage, yang2017optimality}. It is to note that a wrong user ordering leads to an incorrect choice of the power distribution, which may lead to a situation where a few users are always in outage \cite{Ding2014OnthePerformance}. Therefore, it is necessary to acquire the knowledge of the perfect channel gain. The availability of the perfect instantaneous CSI at the transmitter is not a valid assumption in practice, particularly when users of systems like 5G and B5G demand high mobile services. Unpredictable rapid users' movement frequently change channel characteristics that make the perfect CSI estimation process challenging. The error in the perfect channel estimation acts as a source of interference that degrades the overall system performance. The average sum-rate and the outage performance of the NOMA with a perfect CSI are always superior to that of the NOMA with imperfect CSI \cite{Yang2016onthe}. To give a \textit{close to real-time} scenario, researchers consider imperfect CSI models for various network conditions such as slowly varying CSI, delayed CSI feedback, high mobility of users, and so on \cite{Guo2019Robust, Sun2016Cluster, Liang2017Spectrum, Zamani2019Energy}. A second-order statistics (SOS)-based CSI model achieves superior system performance than the imperfect CSI based model \cite{Yang2016onthe}.
 
Another challenging but key-enabling factor of NOMA is the implementation of an error-free SIC decoding. In imperfect SIC, the decoder completely cannot eliminate the signal power of other signals. Consequently, residue power affects the signal detection process as interference in subsequent signal detection. The imperfect SIC not only degrades the overall system performance but also increases processing time due to re-requesting for contents by the users who failed to decode their signals successfully. If a user fails to decode any signals with higher allocated power, it also fails to decode its signal (detailed discussed in section \ref{sec: DecodingSuccessfully}). The imperfect SIC is widely modeled as Gaussian distribution \cite{Kader217Exploiting, Im2019Outage}. However, there may be some typical scenarios where an error does not obey the Gaussian distribution \cite{Li2018Outage}. One interesting fact of NOMA systems is that since the weakest user does not need to apply the SIC process for decoding its signal, there is no effect of imperfect SIC on the performance of the weakest user. A joint CSI- and QoS-based hybrid-SIC process is projected as a promising candidate for next generation multiple access in \cite{liu2022developing}.


\begin{table*}[t h b!]
 \renewcommand{\arraystretch}{1.3}
 \caption{ Comparison between cache-aided NOMA and cache-aided OMA }\label{tab:cacheNOMA_Adv}
  \centering
 \begin{tabular}{| p{22 mm}|  p{149 mm}|   }
   \hline
 \rowcolor{lightgray}
 \centering \textbf{Parameters} & \qquad \qquad \qquad \textbf{Description}\\
\hline
Content popularity & Contents popularity distribution for both the techniques is modeled as Zipf distributed. \\
\hline

 Content selection  & Content selection strategy does not depend on the multiplexing technique (OMA or NOMA).\\
\hline

Cache placement \vspace{1.5 mm}  & \multirow{2}{*}{\parbox{148 mm}{\vspace{1.0 mm}
 \begin{itemize}\setlength{\itemindent}{-0.0 em}
 \vspace{-0.5 mm}
    \item For cache-aided OMA, OMA is used for cache content placement and delivery. 
    \item  For cache aided-NOMA, NOMA is used for cache content placement and delivery.
    \item  Content placement and delivery is faster in cache aided-NOMA than the cache aided-OMA.
\end{itemize}}} \\
\cline{1-1}
Cache delivery  \vspace{1.5 mm} & \\
\hline

System operation  & 
\begin{itemize}\setlength{\itemindent}{-0.0 em}
 \vspace{-2. mm}
    \item If the requested file is cached, the BS sends the file immediately for both techniques. 
    \item If requested file is un-cached, the BS first downloads the requested file from the data center using the backhaul link and then sends it to the users for both techniques. \vspace{-2.0 mm}
\end{itemize}\\

\hline
System performance & 
\begin{itemize}\setlength{\itemindent}{-0.0 em}
 \vspace{-2. mm}
    \item  NOMA always performs better than any conventional OMA when both schemes are equipped with optimal resources \cite{chen2017optimization}.
    \item NOMA serves more users than the OMA under the same network condition \cite{moghimi2020joint}.
    \item The average data rate and the spectral efficiency of NOMA-based system are higher than OMA \cite{rezvani2019cooperative}. \vspace{-1.5 mm}
\end{itemize}\\
\hline 
\end{tabular} 
 \end{table*}
\subsection{Fundamental of Cache-aided NOMA}
 Fundamentally, caching and NOMA are two completely different techniques; caching is a methodology for storing data temporarily in memory (cache), and NOMA is an advanced multiplexing technique for data transmission. Designing a NOMA-enabled system jointly with a cache facility can help both technologies. In a jointly developed cache-aided NOMA system, cache and NOMA individually help each other in their operations and enhance the performance of both techniques. Caching assists NOMA in the interference cancellation during the SIC process and increases the probability of successful decoding. On the other hand, NOMA helps a faster cache placement and delivery process than the OMA. Various studies validated the superiority of cache-aided NOMA over the conventional NOMA system in terms of energy efficiency \cite{Yang2020Cache}, system latency \cite{ fu2019dynamic }, coverage performance \cite{ zhao2018coverage }, spectrum efficiency \cite{ding2017spectral}, successful decoding probability \cite{doan2018optimal}, and outage performance \cite{CodedCache4}. Caching with NOMA becomes an excellent communication technique that can provide support for next-generation communications.
 
 The cache-aided NOMA evolved as an advanced communication concept for next-generation communications. In the OMA technique, users with better channel conditions get higher priority, and users with poorer channel conditions need to wait for access; that initiates fairness and high latency problems. On the other hand, NOMA serves multiple users with various channel gains simultaneously, which provides improved fairness with lower latency \cite {Saito2013}. The performance of NOMA always surpasses any conventional OMA techniques when both are provided with the optimal resource allocation strategies \cite{ chen2017optimization}. The NOMA reaches an enhanced spectral efficiency and power efficiency over OMA \cite{ Sun2017Optimal} \cite{ Wei2017Optimal }. It also achieves an improved sum-rate and individual user rate over the time division multiple access (TDMA) \cite{ Xu2015ANew}. Yang \textit{et al.} validated that the NOMA outperforms traditional OMA in terms of outage probability even when partial CSI is available \cite{ Yang2016onthe}. According to International Mobile Telecommunications (IMT) \cite {series2015imt}, 5G technology needs to support eMBB (requires 100Mbps user data rate), mMTC (needs to provide connectivity to 1 million devices per square kilometer), and URLLC (requires maximum 0.5ms end-to-end latency with reliability above 99.999\% \cite { multiple2016}). The OMA techniques cannot meet the above requirements. On the other hand, the NOMA technique efficiently improves the downlink and uplink spectral-efficiency by 30\% and 100\% respectively in eMBB compared to OMA \cite{ candidate2016}. NOMA-supported mMTC and URLLC applications can serve five and nine times more users, respectively \cite{multiple2016}. The aforementioned advanced features encouraged an amalgamation of the concept of a caching technique with the NOMA over OMA. Table-\ref{tab:cacheNOMA_Adv} compares cache-aided NOMA and cache-aided OMA.

 Due to the dynamic behaviour of the wireless channel and movement of the users, content placement and delivery become challenging in wireless caching networks. NOMA helps in fast content placement and delivery maintaining fairness. A caching strategy need to employ in cache-aided NOMA system to address the following challenges
\begin{itemize}
    \item What and how to cache?
    \item What and when to update?
    \item How to design physical-layer transmission?
\end{itemize}
Superposition coding is widely used for combining signals in the NOMA technique. However, Yaru \textit{et al.} proposed a method for combining signals, named index coding (IC), and claimed that IC is comparably more energy-efficient than superposition coding, particularly when requested files of a pair of users are available in the associated cache \cite{ fu2019mode}. The SIC decoding process of a cache-aided NOMA network is slightly different from conventional NOMA \cite{doan2018optimal}. Notably, caching helps in the decoding process even when the requested file is not cached (detail discussed in Sec \ref{Sec:CIC}). 

 \begin{table*}[t b h!]
 \renewcommand{\arraystretch}{1.3}
 \caption{Cache placement for cache-aided-NOMA systems\label{Cache placement}}
  \centering
 \begin{tabular}{|p{7 mm}|  p{22 mm}| p{13 mm} |p{11 mm}| p{26 mm}| p{32 mm}| p{32 mm}|} 
    \hline \hline 
  \rowcolor{lightgray}
\centering \textbf{Ref. } & \centering \textbf{Objective} & \centering \textbf{System } &\centering \textbf{Caching} & \centering \textbf {Optimization} &  \centering \textbf{Technical Contribution} &  \qquad \textbf{Major Outcome}   \\
\hline

 \cite{ ding2018noma } & {On-peak hour content placement}   & {Edge and D2D caching} & {Uncoded full file} & { Optimization is not employed } &{Proposed two on-peak hour cache content placement strategies} & {Improves cache hit and reduces delivery outage probability}\\
\hline

\cite{ moghimi2020joint } & {Minimize the overall network cost } & {Cooperative edge caching} & {Uncoded full file } & {Hungarian algorithm and successive convex approximation method} &{Propose a novel joint resource allocation and cooperative caching scheme} & {Considerably reduce network cost compared to non-cooperative OMA} \\
\hline

\cite{ li2020cache } & {Minimizes BS average transmit power} & {D2D caching} & {Uncoded full file} & {Quadratic knapsack problem formulation} &{Proposed three methods for cache content placement} &{\textit{Alternating upper plane} method is best for caching} \\
\hline

\cite{rai2020performance} & {Improves delivery delay and sum rate performance}  & {Cooperative edge caching} & {Split file caching} & {Lagrange partial relaxation and McCormick envelopes methods} & {Optimized the cache placement strategy under the constraint of storage capacity} &{Proper resource utilization can achieve ultra-low latency and high throughput} \\
\hline

\cite{yan2019joint} & {Analyzes successful delivery probability
and transmission rate} & {Edge caching} & {Uncoded full file} & {Hierarchical Stakelberg game theory method} & {Proposed joint cache content popularity prediction and access mode selection method} &{Proposed algorithm can achieve the evolutionary equilibrium very fast with 90\% prediction effect} \\
\hline

\cite{Zhao2017ANon} & {Improves spectrum efficiency and reduces outage probability} & {Edge caching} & {Uncoded full file} & {Gale-Shapley algorithm-based distributed method} &{Proposed NOMA-multicasting for pushing and multicasting content simultaneously} & {Proposed scheme is better than conventional OMA-based multicast scheme
} \\
\hline
 
\cite{yin2020qos} & {Increases cache hit and outage performance} & {Edge caching} & {Uncoded full file} & {Optimization is not employed} & {Proposed a QoS-oriented dynamic power allocation strategy} &{Ensures correct detection of far user requested files and improves delay performance} \\
\hline

\cite{Zhang2020Caching} & {Increases cache hit ratio and delivery delay performance}  & {Edge caching} & {Uncoded full file} & {Q-learning based algorithm} &{Proposed long-term caching placement and resource allocation algorithm} & {Trade-off between performance and computational complexity exist} \\
   \hline \hline
   
\end{tabular}
 \end{table*}
 
A few cache placement techniques of NOMA systems has been summarised in Table \ref{Cache placement}. In \cite{ moghimi2020joint}, the authors proposed a cooperative NOMA- caching scheme to analyze the effect of physical storage of BS and available radio resource parameters like QoS, subcarrier assignment, and power allocation constraints in the network cost. The authors also validated that the proposed scheme can improve the network cost reduction compared to other caching strategies and OMA. An optimization problem for content placement in the NOMA system has been formulated to minimize the average transmit power taking into account cache capacity constraints \cite{ li2020cache }. In \cite{rai2020performance}, the authors have optimized the cache placement strategy to reduce the average delay under the constraint of the storage capacity of a fog-computing AP. Xiang \textit{et al.} proposed a coded caching delivery strategy and derived optimal transmit power and rate allocation based on cache status, file sizes, and channel conditions to minimize content delivery latency in cellular networks \cite{xiang2019cache}. A joint cache content popularity prediction and access mode selection problem are formulated as the Stackelberg game in cache-aided NOMA-based F-RANs \cite{yan2019joint}. It is observed that a cache-aided network conventionally stores cache contents during an off-peak time, which may not be an efficient approach, particularly when a  network frequently needs to update the content of the cache. Ding \textit{et al.} proposed NOMA as the best candidate for efficiently storing cache contents during on-peak hours and developed two algorithms, \textit{push-then-deliver strategy} and the \textit{push-and-deliver strategy} \cite{ding2018noma, ding2018application}. The push-then-deliver strategy stores popular content during on-peak times and delivers when requested. The push-and-deliver strategy deals with the scenario when a user's requested content is not cached.


\subsection{Shake hand between NOMA and Cache} \label{shakeHand}
This subsection discusses how NOMA and cache jointly support each other in establishing next-generation communication systems. The caching assists NOMA in the interference cancellation and the successful decoding. On the other hand, NOMA lends a helping hand to caching for placing multiple files within a short free window. 

\subsubsection{Interference Cancellation}
\label{Sec:CIC}
The Cache-enabled interference cancellation (CIC) is an attractive feature of D2D cache-aided NOMA systems. During off-peak time, the BS stores popular content in the cache using the split file caching technique. Perfect knowledge of the cached content is available to the BS. Users get a few segments of the requested file from their associated cache memory and the rest of the portions from the BS. The CIC helps BS to remove a few segments of the requested file from the superimposed signal, which are available in the cache. By identifying the common portions between the received signal and cache contents, the receiver obtains the knowledge of the assigned power to the information associated with those segments. The CIC eliminates the common portions from the superposed signal and reduces the interference power \cite{xiang2019cache, xiang2018cache}. The CIC is a unique feature of the cache-aided NOMA scheme and does not exist in the conventional NOMA.

\begin{figure} [t]
	\begin{center}
		\includegraphics[scale=0.28]{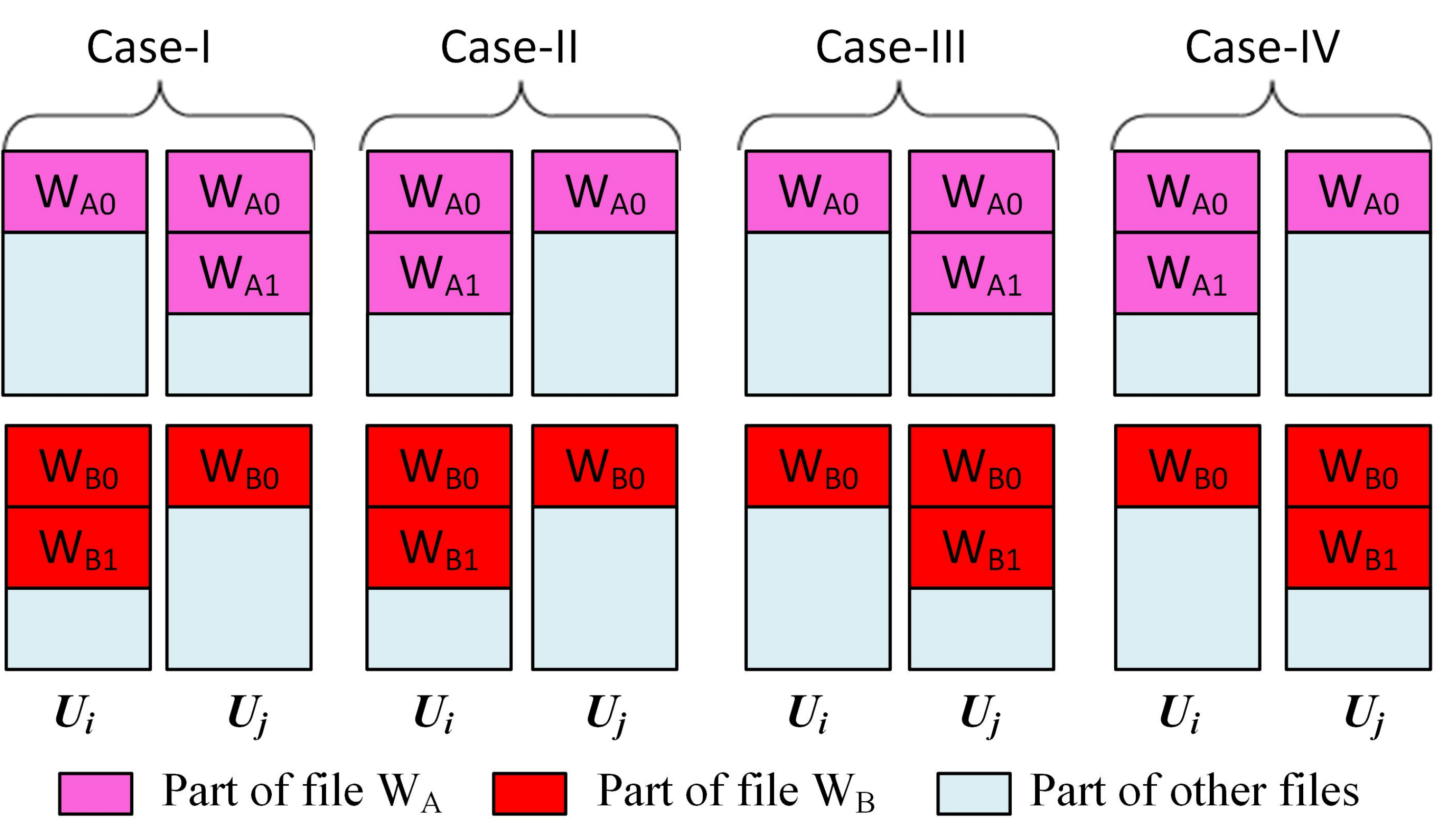}
		\caption{\small Different cache status of $U_i$ and $U_j$ requested file $W_{fn}$, $f \in \{A,B\}$, $n \in \{0,1,2\}$. For all four cases, the file portion $W_{f0}$ and $W_{f2}$ are cached and uncached, respectively. Thus, the transmitter never transmits $W_{f0}$ but always needs to transmit $W_{f2}$ \cite{xiang2019cache}. }
		\label{Fig.CIC Cache}
	\end{center}
	\end{figure}

To understand the CIC, consider a D2D caching system with two cache-enabled users $U_i$ and $U_j$ request for files $W_A$ and $W_B$ respectively. This analysis can be extended for more than two users. The split caching technique is employed, where a file is divided into three segments and placed fully or partially into the cache sequentially to minimize the initial delivery delay. The segments are denoted by $W_{fn}$, $f \in \{A,B\}$, $n \in \{0,1,2\}$. The users cached $c_{kf} \in [0,1]$ portions of $W_f, \; f \in \{A,B\}.$ The minimum and maximum portions of $W_f$ are defined as $\underline{c}_f=min_{k \in \{i, j\}}c_{kf}$ and $\overline{c}_f=max_{k \in \{i, j\}}c_{kf}$ respectively. The corresponding cache status $\underline {k}_f$ = arg $min_{k \in \{i, j\}}c_{kf}$ and $\overline {k}_f$ = arg $max_{k \in \{i, j\}}c_{kf}$. {Four possible cache configurations at the time of the request are shown in Fig.\ref{Fig.CIC Cache}. Case-I is the unfavorable condition for both users, Case-II is favorable for $U_i$ but unfavorable for $U_j$, Case-III is unfavorable for $U_i$ but favorable for $U_j$, and Case-IV is favorable for both users. All these four scenarios can be expressed as follows. \\
\textbf{Case-I:} $i=\overline {k}_B$ and $j=\overline {k}_A$, i.e., $i=\underline {k}_A$ and $j=\underline {k}_B$;\\
\textbf{Case-II:} $i=\overline {k}_B$ and $j=\underline {k}_A$, i.e., $i=\overline {k}_A$ and $j=\underline {k}_B$;\\
\textbf{Case-III:} $i=\underline {k}_B$ and $j=\overline {k}_A$, i.e., $i=\underline {k}_A$ and $j=\overline {k}_B$;\\
\textbf{Case-IV:} $i=\underline {k}_B$ and $j=\underline {k}_A$, i.e., $i=\overline {k}_A$ and $j=\overline {k}_B$;\\
}
	
\begin{figure} [t]
	\begin{center}
		\includegraphics[scale=0.65]{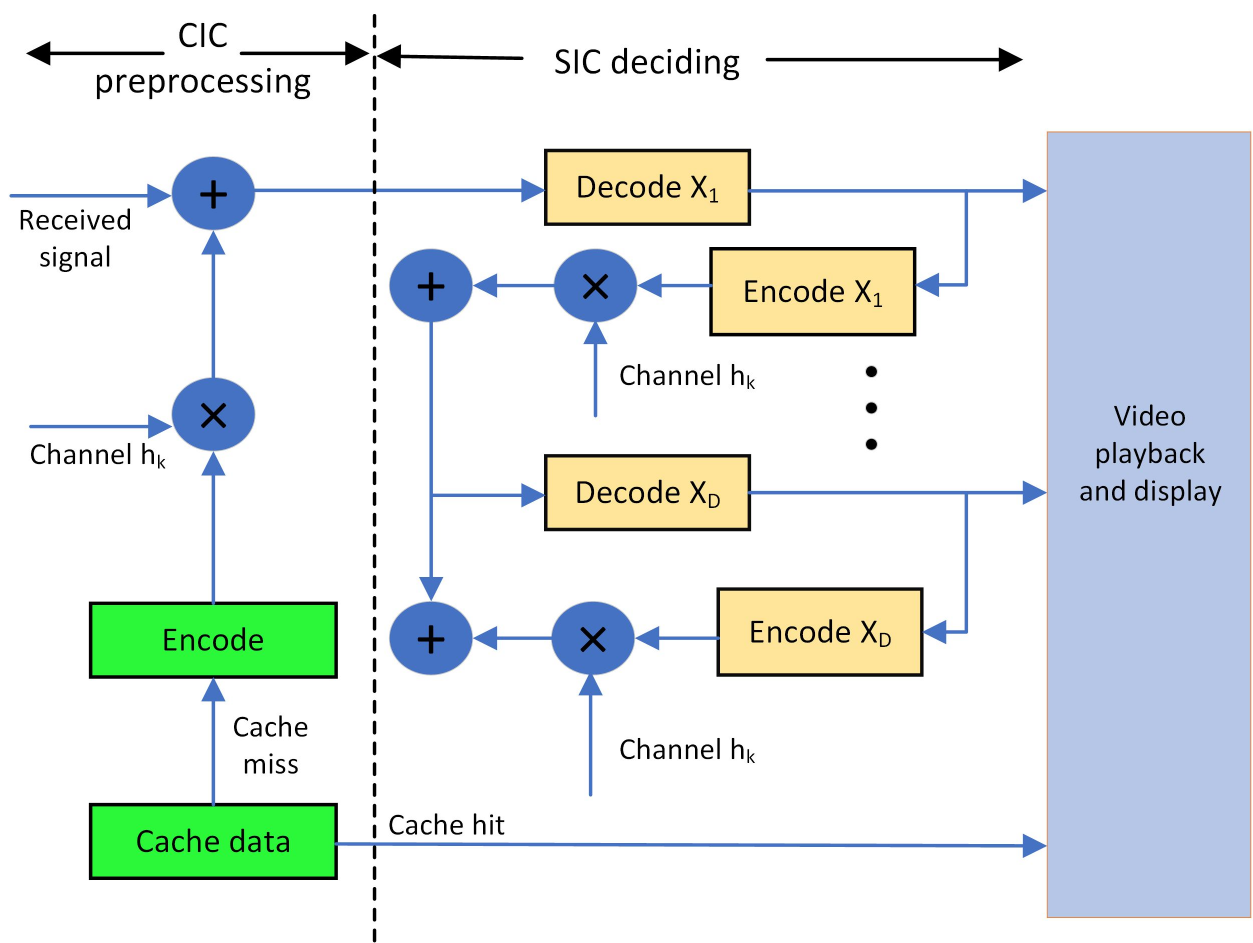}
		\caption{\small Joint CIC and SIC decoding techniques. The cached portions are removed from the received signal by the CIC processing. The residual signals $X_1, X_2,...,X_D $ which are not cached but sequentially decoded by applying traditional SIC process \cite{xiang2019cache}.}
		\label{Fig.CIC}
	\end{center}
	\end{figure}

Due to the limited storage capacity of the cache, placing the whole file $W_A$ or $W_B$ in the cache memory might be impractical. Since the content placement is executed without knowing the actual content demands, caching gain is reduced by full file caching. Therefore, it is assumed that the maximum $W_{f0}$ and $W_{f1}$, $f \in \{ A, B\}$ can be cached. The BS knows the perfect cache statuses, thus, transmits only the un-cached subfiles.The BS transmits $x$, a superimposed NOMA signal given by 

\begin{eqnarray}
{x}=
\begin{cases}
\sqrt{p_{i,1}}x_{A1}+\sqrt{p_{i,2}}x_{A2}+\sqrt{p_{j,1}}x_{B1}\\
\qquad \qquad \qquad \qquad \quad \;+\sqrt{p_{j,2}}x_{B2}, \;\quad \text{Case-I},\\
 \sqrt{p_{i,2}}x_{A2}+\sqrt{p_{j,1}}x_{B1}+\sqrt{p_{j,2}}x_{B2}, \quad  \text{Case-II},\\
 \sqrt{p_{i,1}}x_{A1}+\sqrt{p_{i,2}}x_{A2}+\sqrt{p_{j,2}}x_{B2}, \quad   \text{Case-III},\\
 \sqrt{p_{i,2}}x_{A2}+\sqrt{p_{j,2}}x_{B2}, \qquad \qquad  \qquad \; \text{Case-IV},
 \end{cases}\
\label{eq: CIC_1}
\end{eqnarray}
 where $x_{fs}$ is the codeword corresponding to un-cached subfile $W_{fn}$, $f \in \{A,B\}$, $s \in \{1,2\}$, and $p_{kn}$, $k \in \{i,j\}$, $s \in \{1,2\}$ is the transmit power to $x_{fs}$. The joint CIC and SIC receiver, shown in Fig. \ref{Fig.CIC} performs the CIC pre-processing before applying the SIC-based decoding. The CIC process exploits the cache of $U_i$ to remove $X_{B1}$ and $X_{A1}$ form Case-I and Case-II respectively of (\ref{eq: CIC_1}), and we get residual signal (\ref{eq: CIC_2}). Similarly, the CIC process exploits the corresponding cache of $U_j$ to get (\ref{eq: CIC_3}) from (\ref{eq: CIC_1}). 
  
 \begin{eqnarray}
{y^{CIC}_i}=
\begin{cases}
h_i\left(\sqrt{p_{i,1}}x_{A1}+\sqrt{p_{i,2}}x_{A2}+\sqrt{p_{j,2}}x_{B2}\right)\\
\qquad \qquad \qquad \qquad  +z_i, \;  \text{Case-I \& III},\\
 h_i\left(\sqrt{p_{i,2}}x_{A2}+\sqrt{p_{j,2}}x_{B2}\right)\\
\qquad \qquad \qquad \qquad +z_i, \;  \text{Case-II \& IV}.
 \end{cases}\
\label{eq: CIC_2}
\end{eqnarray}

\begin{eqnarray}
{y^{CIC}_j}=
\begin{cases}
h_j\left(\sqrt{p_{i,2}}x_{A2}+\sqrt{p_{j,1}}x_{B1}+\sqrt{p_{j,2}}x_{B2}\right)\\
\qquad \qquad \qquad \qquad  \; +z_j, \; \text{Case-I \& II},\\
 h_j\left(\sqrt{p_{i,2}}x_{A2}+\sqrt{p_{j,2}}x_{B2}\right) \\ \qquad \qquad \qquad \qquad \; +z_j, \;  \text{Case-III \& IV}.
 \end{cases}\
\label{eq: CIC_3}
\end{eqnarray}

The CIC process discards the file segments (all/a few) belonging to other users if those are available in the associated cache. After the CIC operation, the receiver applies the conventional SIC process to decode its signal. It is noted that before applying the SIC process, a receiver can remove a few amounts of interference from the received superimposed signal by using the CIC process. The caching technique helps SIC decoding even when the requested file is not cached. The combination of caching and NOMA can only provide this advantage. The joint CIC and SIC decoding process significantly increases the sum-rate and reduces the file delivery times \cite {xiang2019cache, xiang2018cache }. In \cite{shen2020cache}, the authors have proposed a new D2D cache-aided NOMA system, where the cache infrastructure is available at both the users’ end and the BS, and employing CIC enhances the system sum-rate. The caching technique also is implemented in MIMO-based wireless networks for canceling interference \cite{ Wei2019Cache}.

\subsubsection{Probability of Successful Decoding}
\label{sec: DecodingSuccessfully}
\begin{figure} [ t b!]
	\begin{center}
		\includegraphics[scale=0.75]{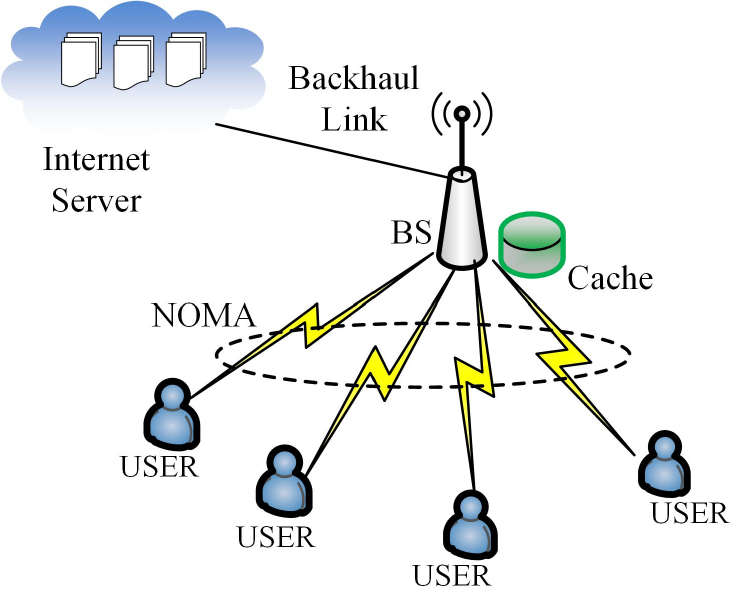}
		\caption{\small Basic cache-aided NOMA network, where BS serves multiple users simultaneously other than time and frequency domain, and cache reduces use of backhaul link.}
		\label{Fig.CacheNomaSystem}
	\end{center}
\end{figure}

This section discusses how full file edge caching helps NOMA in the SIC-based decoding process. Unlike split file caching, BS stores complete files in the cache for full file caching. In NOMA, receivers applies SIC process to subtract a large number of signals intended for other users from the superposition signal before decoding its signal. A decoding failure occurs when a user fails not only to decode its signal but also anyone of the \textit{other signals}. In a cache-aided NOMA system, a user needs to decode comparably a smaller number of \textit{other signals} as contents of some users may be available in the cache. Consequently, a cache-aided NOMA attains an enormous improvement in SDP. To understand the decoding process, consider a simple edge cache-aided NOMA system with $K$ users, $\mathcal{K} \in \{1,2,...,K\}$ and a cache-enabled BS, as shown in Fig.\ref{Fig.CacheNomaSystem}. The channel coefficient of BS-to-\textit{k}th user is $h_k$.
Let $u_1$ and $u_2$ request for files $f_1$ and $f_2$ respectively to the BS. Now, four possible scenarios of cache status are:\\
\textit{Scenario-1}: Both the $f_1$ and $f_2$ are cached\\
\textit{Scenario-2}: $f_1$ is cached but $f_2$ is not cached\\
\textit{Scenario-3}: $f_1$ is not cached but $f_2$ is cached\\
\textit{Scenario-4}: Both the $f_1$ and $f_2$ are not cached

According to NOMA principles, the transmitter allocates the maximum power to the user having the poorest channel gain (weakest user) and minimum power to the user with the strongest channel gain (strongest user). The weakest user decodes its signal directly considering other signals as interference. The stronger users apply the SIC process during decoding their signals. Our aim is not to derive the SDP of a cache-aided NOMA network rather to illustrate the impact of caching on the decoding process. Hence, for simplicity, we assumed $|h_1|^2 < |h_2|^2$, i.e., $u_1$ is weaker than the $u_2$ therefore, BS allocates higher power to $u_1$ than the $u_2$.

\textit{Scenario-1}: In this case, both $f_1$ and $f_2$ are available in the cache. Let $\mathcal{S}_1^{(1)}$ be the SINR of the signal for $u_1$ at $u_1$. Being a stronger user, $u_2$ first decodes the signal of $u_1$ and then decodes its signal using SIC. Consider, $\mathcal{S}_1^{(2)}$ and $\mathcal{S}_2^{(2)}$ are the SINRs of the $u_1$'s signal and $u_2$'s signal, respectively, at $u_2$. Now, $\mathcal{D}^{(1)}$, the overall SDP of the system for scenario-1 can be given as
\begin{equation}
\label{eq: Scenario-1}
    \mathcal{D}^{(1)} = \mathcal{P}_r \bigg(\mathcal{S}_1^{(1)} \ge \gamma \bigg) \mathcal{P}_r \bigg (min \big( \mathcal{S}_1^{(2)}, \mathcal{S}_2^{(2)} \big) \ge \gamma \bigg ),
\end{equation}
where $\gamma$ is the predefined threshold SNR required to decode signal successfully. The first term $\mathcal{P}_r(\mathcal{S}_1^{(1)} \ge \gamma )$ and the second term $\mathcal{P}_r(min ( \mathcal{S}_1^{(2)}, \mathcal{S}_2^{(2)} ) \ge \gamma )$ of (\ref{eq: Scenario-1}) are the SDP of  $f_1$ and $f_2$ respectively.

\textit{Scenario-2}: In this scenario, $f_2$ is not cached. BS needs to access the internet server for $f_2$ through backhaul link. Let $\mathcal{S}_{BS}^{(2)}$ is the SNR of the $u_2$'s signal at BS. Once the BS receives $f_2$ from internet server it delivers both the $f_1$ and $f_2$ using NOMA. Now, $\mathcal{D}^{(2)}$, the overall SDP for the scenario-2 can be expressed as

\begin{equation}
    \mathcal{D}^{(2)}= \mathcal{P}_r\bigg(\mathcal{S}_1^{(1)} \ge \gamma \bigg ) \mathcal{P}_r \bigg(min \big ( \mathcal{S}_1^{(2)}, \mathcal{S}_2^{(2)}, \mathcal{S}_{BS}^{(2)} \big ) \ge \gamma \bigg).
    \label{Eq.Decode2}
\end{equation}
 
\textit{Scenario-3}: In this scenario, $f_1$ is not cached. Similar as scenario-2, after receiving $f_1$ from internet server, BS delivers both the files to the users. The SNR of the  $u_1$'s signal at BS is $\mathcal{S}_{BS}^{(1)}$. $\mathcal{D}^{(3)}$, the overall SDP for the scenario-3 is given by

\begin{equation}
    \mathcal{D}^{(3)} = \mathcal{P}_r \bigg (min \big ( \mathcal{S}_1^{(1)}, \mathcal{S}_{BS}^{(1)} \big ) \ge \gamma \bigg ) \mathcal{P}_r \bigg (min \big ( \mathcal{S}_1^{(2)}, \mathcal{S}_2^{(2)} \big ) \ge \gamma \bigg  ).
    \label{Eq.Decode3}
\end{equation}

\textit{Scenario-4}: This is the case when none of the file is cached. BS downloads both the $f_1$ and $f_2$ from internet server using NOMA. It is assumed that internet server allocates more power to the $f_1$. After receiving both the files $f_1$ and $f_2$ from internet server, BS decodes the files and then delivers both the files to corresponding the users using NOMA. Now, $\mathcal{D}^{(4)}$, the overall SDP is expressed as
\begin{equation}
\begin{split}
\mathcal{D}^{(4)} &= \mathcal{P}_r \bigg (min \big ( \mathcal{S}_1^{(1)}, \mathcal{S}_{BS}^{(1)} \big ) \ge \gamma \bigg )\\
  & \qquad  \mathcal{P}_r \bigg (min \big ( \mathcal{S}_1^{(2)}, \mathcal{S}_2^{(2)}, \mathcal{S}_{BS}^{(1)}, \mathcal{S}_{BS}^{(2)} \big ) \ge \gamma \bigg  ).
    \label{Eq.Decode4}
    \end{split}
\end{equation}

The first term and second term of (\ref{eq: Scenario-1})-(\ref{Eq.Decode4}), are the SDP of $f_1$ and $f_2$ respectively. It is worth noting that without caching (Scenario-4), multiple users need to decode a sequence of signals intended for other users before obtaining their signals. On the contrary, users can discard signals of others users using the cached content, which improves the SDP.


\subsubsection{Cache Placement Using NOMA}
 The 5G and B5G need to share the spectrum among UEs to improve the spectral efficiency and simultaneously ensure massive connectivity with a high reliability. The cache placement during off-peak hours is not an efficient approach for 5G. In comparison with the OMA technique, the NOMA can place more content in the cache and serve a large number of users during the content delivery phase within a short duration of time. The OMA can store (or push) only a single file during a single time slot. Therefore, the BS pushes only the content with a maximum popularity during the first time slot and the second most popular file during the second time slot, and so on. When a comparatively longer period is available, the OMA-based content placement requires sophisticated methods which efficiently schedule the files based on their popularity \cite{ding2018noma}. Unlike the OMA, during a single time slot, applying the NOMA principle, the BS can push multiple files based on their popularity at the same time. The content delivery phase is divided into small time slots. During a single window, OMA can serve a single user whose requested file is available in the cache. An efficient user scheduling algorithm based on \textit{first-in-first serve} is required for serving multiple users' requests. On the other hand, like the content pushing phase, the NOMA serves multiple users simultaneously. The cache-aided NOMA scheme efficiently improves the cache hit probability and reduces the delivery outage probability compared to conventional OMA-based caching.

\textbf{Summary:}  Various studies validated that under the optimal scenario for both NOMA and caching, NOMA always outperforms OMA \cite{ chen2017optimization}. The performance of NOMA-based systems crucially depends on the degree of accuracy of the SIC process. NOMA with caching can remove interference fully or partially using cached contents and increase the SDP. Cache and NOMA individually help each other. The NOMA makes a faster cache placement process and delivers requested files simultaneously to multiple users maintaining fairness. On the other hand, caching helps NOMA in the SIC decoding process even when requested files are not cached. The conventional cache strategies push contents during off-peak hours, which is not an efficient approach, specifically when the popularity of contents changes suddenly and networks need to update cache contents. NOMA-aided caching can push multiple contents within a short duration and becomes the best candidate for content pushing during peak time \cite{ding2018noma}. Push-then-deliver and the push-and-deliver strategies are capable of content placement during on-peak hours \cite{ding2018noma}. A combination of NOMA with caching could be a breakthrough strategy for next-generation communication systems.

\begin{table*}[t h!]
 
 \renewcommand{\arraystretch}{1.3}
 \caption{Different performance metrics and their solution techniques}\label{Tab:performance Metrric}
  \centering
 \begin{tabular}{|p{8mm}|p{10 mm} | p{22 mm}| p{44 mm}| p{70 mm}|  }
    \hline \hline
  \rowcolor{lightgray}
\centering \textbf{Ref. } &\centering \textbf{System} &\centering \textbf{Metrics}  & \centering \textbf{Implemented Technique} &  \qquad \qquad \qquad \textbf{Research Gap/ Merit} \\
\hline
\cite{ shen2020cache } & {D2D} & {sum-rate} & {CIC process} & {Assumed perfect knowledge of CSI and error free SIC process} \\
\hline
\cite{zhao2018resource} & {D2D} & {sum-rate}  & {Subchannel allocation} & {Considered equal power allocation, contents are of equal size}\\
\hline
\cite{ pei2020hybrid } & {ECT} & {sum-rate} & {Non-convex optimization problem } & {Imperfect CSI is taking into account} \\
\hline
\cite{ xiang2019cache } & {D2D} & { Sum-rate \& delay minimizing} & {Optimal decoding order and optimal transmit power } & { Proposed delivery scheme is applicable for any caching scheme} \\
\hline
\cite{ fu2021efficient  } & {ECT} & {Delay
minimizing } & {Recommendation Mechanism} & {Perfect CSI is known at BS} \\
\hline
\cite{ fu2019dynamic } & {ECT} & {Delay
minimizing} & {Deep neural network-based dynamic power control} & {Assumed guaranteed successful decoding at receivers} \\
\hline
\cite{Liu-Yu18}  & {ETC} & {Delay
minimizing } & {Resource allocation approach} & {Considered a typical static single user scenario} \\
\hline
\cite{ li2019resource} & { ETC} & {Delay minimizing}  & {Formulated optimization problem} & {Considered an equal file size}   \\
\hline
\cite{ doan2018optimal  } & {D2D} & {SDP} & {Optimal power allocation}& {Considered two-user scenario with perfect CSI} \\
\hline
\cite{gurugopinath2019cache, mohan2020cache } & {D2D} & {SDP} & {Power allocation optimization} & {Considered a perfect SIC technique} \\
\hline
\cite{ doan2019power } & {D2D } & {SDP} & {Deep learning-based power allocation} & { Considered scheduling delay of content requested by users} \\
\hline
 
 \cite{yin2020qos} & {ECT} & {Outage and cache hit probability} & {Dynamic power allocation strategy}  &  {BS cannot serve users during the periodic cache placement process}  \\
\hline

 \cite{zheng2020outage} & {D2D} & {Outage performance} & {Inverse Laplace transform used for exact outage probabilities} & {Evaluated expression for exact outage probabilities} \\
 \hline
 
\cite{ dani2021noma } & {D2D} & {Outage performance} & {Power allocation and user pairing scheme} &  {Hybrid delivery scheme can select NOMA or coded multicasting based on the channel conditions.}  \\

   \hline \hline
\end{tabular}
 \end{table*}

\section{Key  Performance  Indicators}\label{sec: KPI}
This section presents a fundamental analysis of key performance indicators (KPIs) of cache-aided NOMA-based 5G and B5G systems. Various Performance metrics and the approaches adopted to enhance them have been summarised in Table-\ref{Tab:performance Metrric}.

\subsection{Sum-Rate Maximization}
Sum-Rate measures the successful data transmission rate over the communication channel of the unit bandwidth. Achieving a higher sum-rate is a primary requirement for any communication system especially when a video file is streaming. The information-theoretic studies have demonstrated that NOMA cannot elevate the overall sum capacity of the system compared to conventional OMA \cite{tse2005fundamentals, el2011network}. Hence, NOMA is exploited to maintain user fairness \cite{Ding17, Ding2017Asurvey, Sun2017Optimal, Hanif2016A}. Ding \textit{et al.} demonstrated that fixed power allocation based NOMA system achieves remarkable throughput gain only for asymmetric channel gain quality of the users, and for symmetric channel gains performance of NOMA and OMA are identical \cite{ding2015impact}. Wireless caching is an efficient approach incorporated with the NOMA to increase the sum-tare of 5G and B5G networks \cite{wong2017key}. The cache-aided NOMA significantly increases the achievable sum-rate compared to OMA by enabling joint CIC and SIC \cite{ xiang2019cache }. In \cite{ zhao2018resource}, authors validated that cache-aided cloud radio access networks can achieve an improved sum-rate when NOMA is incorporated. The authors proposed a cache-aided NOMA-based D2D system, where a pair of users utilize the uplink channels for delivering cached contents \cite{shen2020cache }. The performance of the proposed network paradigm was evaluated in terms of the sum-rate. Xinyue \textit{et al.} \cite{ pei2020hybrid } formulated a sum-rate maximization problem under the constraints of the peak allocated power, backhaul capacity, minimum unicast rate, and maximum multicast outage probability for evaluating the performance of a cache-aided NOMA-based multiple-input single-output system.

\subsection{Delay Reduction}
The delay in delivering requested files depends on various network resources like transmit power, available bandwidth, cache status, etc. Overall transmission delay of a cache-aided NOMA system depends on the delay associated with both the backhaul and BS-to-user links. Under a cache hit condition, BS delivers requested files from the cache and reduces the delivery delay by saving the time required to fetch the requested file from the data center using the backhaul link. It is another advantage of cache when incorporated into a NOMA technique. To analyse the delay reduction technique, consider a cache-aided NOMA network, as shown in Fig.\ref{Fig.CacheNomaSystem}, where \textit{k}th user requests for \textit{i}th $\forall i \in \mathcal{I}=\{1,2,...,I \} $ file of size $L_i$, $i \in \mathcal{I}$ to the BS. Transmission over the backhaul link is subjected to the availability of the requested file in the cache. The cache status of the \textit{i}th content is symbolized as $C_i \in \{0,1\}$. Particularly, $C_i=1$ if the requested content is cached and $0$ otherwise. Assuming $R_B$ as the transmission rate of the backhaul link, $\mathcal{T}_B$, transmission time required for \textit{i}th file is given by
\begin{equation}
    \mathcal{T}_B= (1-C_i)\frac{L_i}{R_B}.
\end{equation} 

The cache-aided NOMA system reduces the delivery delay by $\mathcal{T}_B={L_i}/{R_B}$ when requested file is cached. The dynamic power allocation plays a vital role in reducing the delivery delay of cache-aided NOMA systems, where the volume of data is different \cite{fu2019dynamic }. In \cite{fu2019dynamic}, for a cache-aided NOMA system, authors minimize the data transmission delay of each user under the constraints of the total available power and maximum tolerable transmission delay. Liu \textit{et al.} performed joint tasks scheduling and resource management to minimize the transmission latency subjected to maximum transmission delay and network cost for a cache-enabled ultra-dense network \cite{Liu-Yu18}. The delivery delay of a cache-aided NOMA is minimized by jointly optimizing the decoding order of NOMA and power and rate allocations \cite{xiang2018cache}. In \cite{ kim2019grouped}, the authors have jointly optimized the transmission strategies for both backhaul and BS-to-users links under the constraint of transmission power to minimize the delivery time of an edge caching-enabled NOMA system. Recently, in \cite{ fu2021efficient}, the average transmission delay of a cache-aided NOMA network with two-user is derived considering a scenario when both users request files of the same size. The average delay under the constraint of the minimum quality requirement of the file is minimized considering a \textit{recommendation} mechanism \cite{ fu2021efficient}. 
 
\subsection{Energy Consumption Minimization}
Reducing the energy consumption is crucial for communication systems specifically battery operated systems. Various energy consumption minimization approaches adopted in cache-aided NOMA networks. The authors reduce the average signal transmitted power of a cache-aided NOMA-based cellular network under the constraint of cache memory capacity \cite{li2020cache}. In \cite{Yang2020Cache}, jointly optimizing the task offloading, computation and cache resource allocation under the constraint of caching and computing resources, authors have minimized the total energy consumption for the proposed cache-aided MEC network. In \cite{fu2018optimal}, authors minimized the total required transmitted power subjected to a minimum data rate of the users. Increasing energy efficiency is one of the challenging issues of Unmanned Aerial Vehicle (UAV)-assisted wireless networks. The energy efficiency of a UAV-assisted wireless NOMA system is maximized under the constraint of subchannel assignment and power allocation to cache-enabled UAVs \cite{ li2020subchannel}. In \cite{ li2019resource}, authors propose a two-sided matching and swapping algorithm  for maximizing the energy efficiency of a UAV-assisted NOMA-based fog wireless network. Authors are aiming to reduce the total consumption of energy by the UAV-assisted NOMA-based MEC networks taking into account the task computation allocation, computation capacity, and UAV trajectory in \cite{ budhiraja2021energy}. The authors in \cite{ wen2019interference } studied resource allocation for enhancing the energy efficiency by optimizing subchannel allocation and power allocation in a NOMA hierarchical network. 

\subsection{Successful decoding probability} Successful decoding probability (SDP) defines the probability that a receiver decodes own signal correctly. Cache-aided NOMA systems achieves an improved SDP than NOMA and cache-aided OMA systems \cite{ doan2018optimal, mohan2020cache}. The optimal power distribution is the most popular approach for enriching the SDP. The authors formulated optimal power distribution problems over the Rayleigh fading channel \cite{doan2018optimal}, Weibull, Nakagami-m, and Rician downlink fading channels in a cache-aided cellular network \cite{ mohan2020cache}. Depending on the QoS, Yin \textit{et al.} proposed a dynamic power allocation strategy for a cache-aided cellular system that increases the probability of successfully decoding compared with the OMA and fixed power allocation-based NOMA schemes \cite{yin2020qos}. Doan \textit{et al.} in \cite{doan2019power} propose divide-and-conquer-based and deep-learning-based power allocation methods to maximize the SDP that also ensures QoS and fairness of the users.

Apart from the above-mentioned parameters, cache hit probability, backhaul cost, outage probability, network delay, spectral efficiency, and energy efficiency are KPIs for designing cache-aided NOMA systems. 

\textit{Outage Probability-} Outage probability and SDP are the two sides of the same coin. Outage probability tells the probability that a receiver fails to achieve QoS above the threshold level. On the other hand, SDP conveys the probability that a receiver successfully decodes its signal, i.e., QoS is above the threshold level. Outage probability also a widely used to evaluate the performance of  cache-aided systems. Dani et al. evaluate the performance of the proposed hybrid NOMA or coded multicasting-based delivery scheme in terms of outage probability \cite{ dani2021noma, CodedCache4 }. These articles validate the superiority of NOMA over coded multicasting when channel gains of the paired users are highly distinctive, and coded multicasting is better than NOMA under similar channel gains. The outage performance of a MIMO NOMA cellular network is analyzed using inverse Laplace transform in \cite{zheng2020outage}. 

\textit{Cache Hit Probability-} The hit rate is an important parameter that tells how professionally popular files are selected to place in the cache. It is a ratio of the number of times the cache successfully delivers requested files and the total number of times users request files. The hit rate primarily   evaluates the efficiency of cache placement techniques. A successful cache hit reduces communication costs by avoiding the use of the backhaul link and also reduces outage probability significantly \cite{yin2020qos}. The performance gain margin of cache-aided NOMA systems over the conventional NOMA systems crucially depends on the higher cache hit probability. 

\textit{Backhaul Cost}- BS communicates with other BSs or internet servers via a backhaul link. Generally, expensive optical fibers are used as a backhaul link to achieve high-speed data transfer \cite{li2018survey}. Inefficient utilization of backhaul leads to an increased overall expenditure of wireless communication systems. The cache technique efficiently reduces the Backhaul cost. The computation time, successful cache hit, and content delivery latency play significant roles in increasing the energy efficiency of cache-aided wireless networks.  

\textbf{Summary:} This section discussed some vital performance metrics of 5G and B5G systems. Table \ref{Tab:performance Metrric} shows interesting insights into caching techniques applied to evaluate the performance of cache-aided NOMA systems. The D2D caching techniques are widely implemented to maximize the SDP and minimize the outage probability. On the other hand, the edge caching has been implemented to reduce the delivery delay. The performance improvement of cache-aided NOMA systems in terms of any metric depends on the higher cache hit probability.


\begin{figure*} [ h t!]
	\begin{center}
		\includegraphics[scale=0.61]{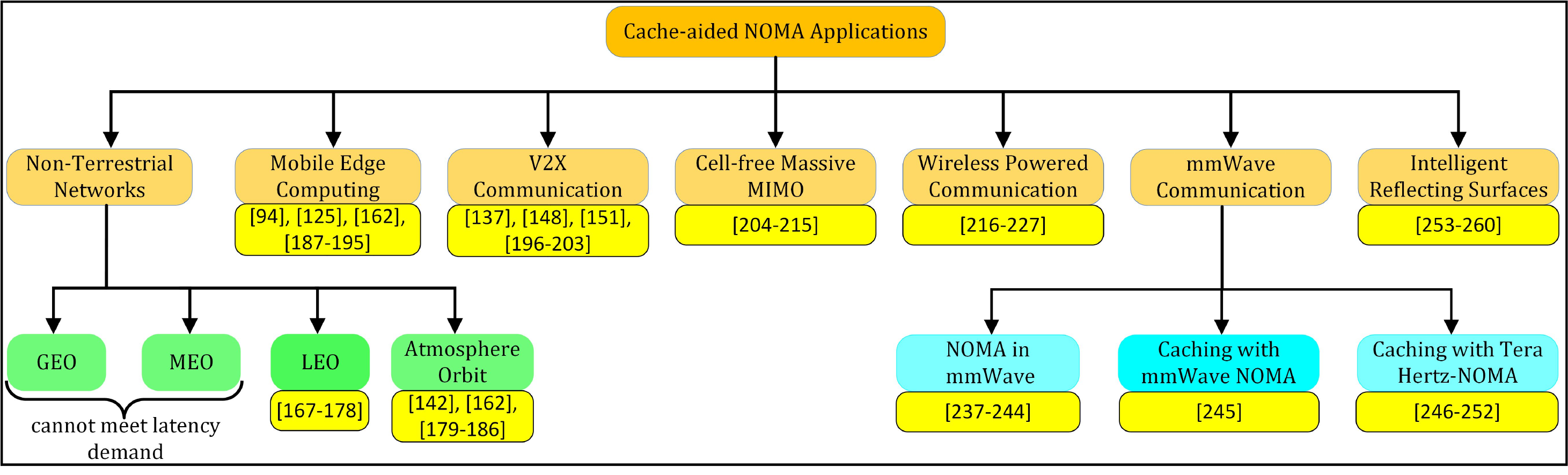}
		\caption{\small Application areas of the cache-aided NOMA technique.}
		\label{Fig.Applications}
	\end{center}
\end{figure*}

\section{Verticals and Use Cases}\label{sec: application}
This section focuses on various application scenarios of cache-aided NOMA networks and the challenges associated with these applications along with their solutions. The application scenarios of cache-aided NOMA are presented in Fig.\ref{Fig.Applications}.

\subsection{Non-Terrestrial Networks}
Non-terrestrial networks (NTN) have evolved a lot over the last decade, beyond simple drones, and they now encompass a whole eco-system like hierarchy. The drones or unmanned aerial vehicles (UAVs) are at the lowest layer of the atmosphere superseded by cubesats in the second layer at the edge of atmosphere, the low-earth-orbit (LEO) satellites form the third layer beyond the atmosphere, and finally, the geostationary-earth-orbit (GEO) satellites are in the topmost layer in deep space. The backhaul link of terrestrial communication networks is connected with the data center via an optical fibers link which is prone to damage during disasters. It is challenging to deploy terrestrial infrastructures in remote locations like mountains, seas, and deserts. Moreover, the speed of light in optical fiber is 30-40\% slower than that of free space, which has motivated researchers for non-terrestrial communications. A few companies like Google, Facebook have already launched satellites for providing better QoS with low latency to their customers. The non-terrestrial network can cover a large area on Earth for an instant, a LEO satellite approximately covers $1$ million km$^2$ area. Various airborne platforms such as Balloons \cite{Balloon2017}, Helikites \cite {Chandrasekharan2016Designing}, and UAVs \cite{ Zeng2016Wireless} are recently emerging as potential approaches to meet wireless traffic demands, specifically for mobile users. Based on the orbital altitude, the non-terrestrial networks are classified into four categories,  (i) GEO (35786 km), (ii) Medium Earth Orbit (MEO) (2000 – 35000 km), (iii) LEO (160 – 2000 km), and (iv) Atmosphere Orbit (a few hundred meters).
\begin{figure} [t]
	\begin{center}
		\includegraphics[scale=0.5]{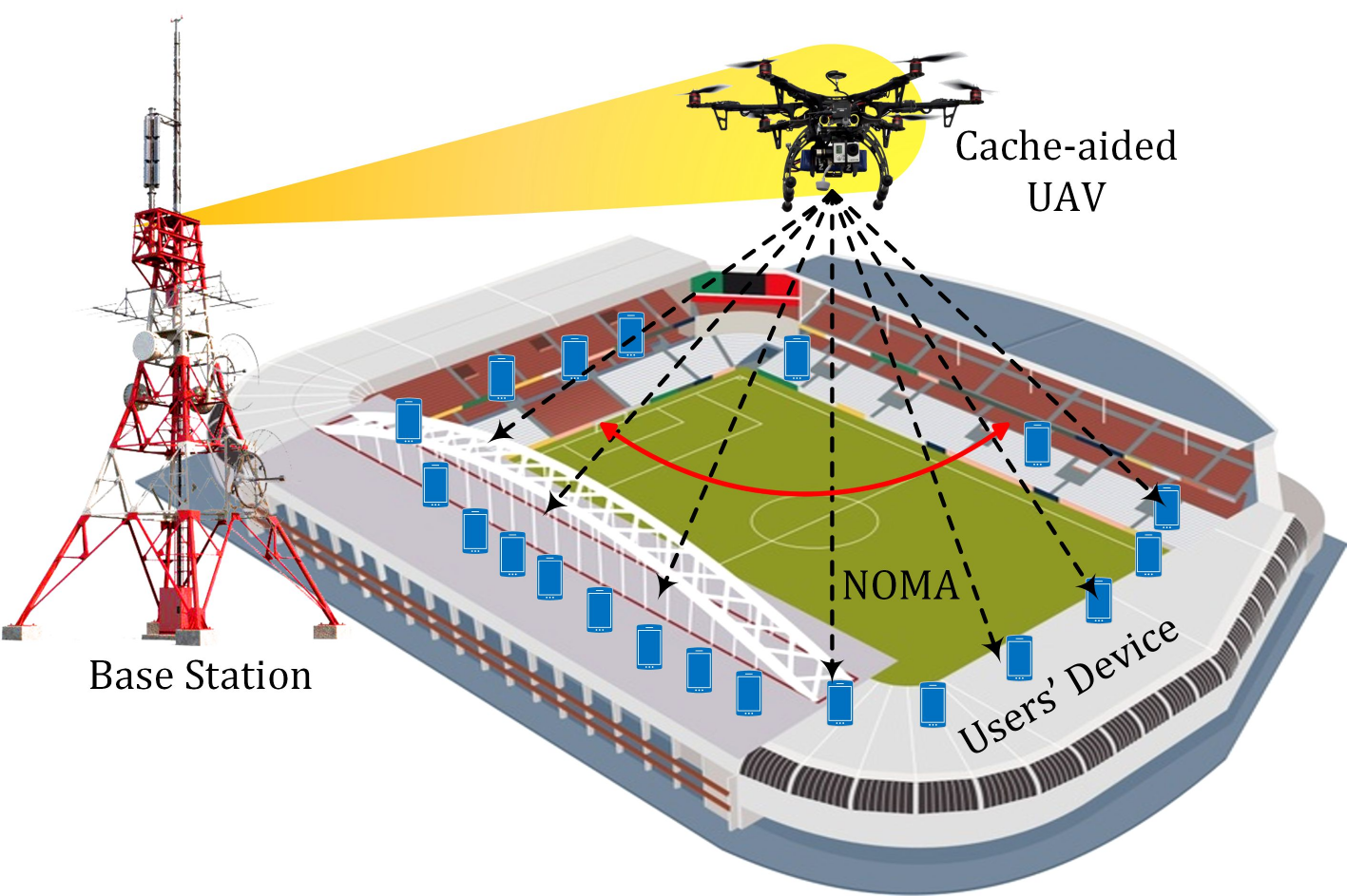}
		\caption{\small One of the popular application scenarios of cache-aided NOMA-based UAV-assisted communication systems. The UAV was deployed temporarily to assist the BS to meet thousands of users' requests \cite{Zhang2020Caching}}
		\label{Fig.Application_UAV}
	\end{center}
	\end{figure}

A signal requires a much higher round trip time between ground station and satellites deployed in MEO (or above), thus cannot meet the latency demand for 5G communications. In LEO satellite-based communications (LEOComm), as the satellite is deployed in relatively lower altitude orbits, the round trip time of a signal is only a few ms (10-15ms for the SpaceX Starlink system) \cite{ SU2019Broadband } and can meet the latency requirement of Internet of Things (IoT), smart grid, and vehicular communication applications \cite{ Hassan2020Dense}. The caching facility in the satellite networks improves the latency performance and makes LEOComm a possible candidate for the 5G paradigms. Armon \textit{et al.} have designed and addressed operating issues of cache-aided satellite distribution systems for web caching \cite{ Armon2004Cache, Armon2003Cache }. To minimize the downlink and uplink traffic load, Wu \textit{et al.} have proposed a two-layer caching model for satellite-terrestrial networks, where caching in the ground stations constitutes the first layer and caching in the satellite constitutes the second layer \cite{ Wu2016Two, Ngo2021Two}. The authors have validated that two content caching schemes, named as \textit{most popular content-based} and \textit{uniform content-based} schemes can efficiently improve the spectral efficiency in the Hybrid satellite-terrestrial relay networks \cite{ An2019Onthe }. Google Loon project is one of the industry projects where Google have installed Internet-delivery drone for providing global massive connectivity \cite{ Facebook2014}. CubeSats are a class of miniaturized satellite for research build up by multiple cubic modules of dimensions 10 cm $\times$ 10 cm $\times$ 10 cm deployed into the lower altitude of LEO. It is reported that $1634$ CubeSats already launched by Aug. 2021, and the future of nanosatellites is still to come \cite{KuluNano}.
Researchers are actively investigating the applicability of cache-aided NOMA in hybrid satellite-terrestrial (HST) networks to improve spectral efficiency, outage probability, hit probability and transmission latency \cite{zhang2020performance, zhang2021noma, singh2022performance}. In \cite{zhang2020performance}, Zhang \textit{et al.} analyzed the transmission delay and cache hit probability of cache-aided NOMA HST systems, where users receive their requested contents from a cache-enabled relay node.  In addition, they also investigated the outage probability and hit probability of a cache-aided NOMA HST network \cite{zhang2021noma}. Recently, Vibhum \textit{et. al} have incorporated a cache-aided NOMA in overlay-based cognitive hybrid satellite-terrestrial networks, where a secondary transmitter with cache capability is employed for cooperative relaying following the NOMA protocol \cite{singh2022performance}. Unlike \cite{zhang2020performance}\cite{zhang2021noma}, users receive signals from the relay and the satellite directly in \cite{singh2022performance}. Herein, authors validated the superiority of the cache-aided NOMA over the cache-free NOMA model in terms of the outage probability. The cache-aided NOMA technique has not been explored much in satellite networks and could be a promising research domain.  
 
UAVs deployed in the atmospheric orbit are the most popular and efficient commercial approaches to provide short-term connectivity in a hot-spot area. UAVs-aided wireless communication is gaining attention among researchers from both the industrial and academic communities for its low infrastructural cost, reduced size, line-of-sight communications, and flexible deployment process. Though UAV was developed for military applications but presently is being utilized for commercial applications also. To fulfill the rising demand for high data transmission rate with low latency, UAV exploited as an effective approach for highly dense wireless communication networks \cite{Mozaffari2017Wireless, Cui2020Multi}. The wireless systems deploy UAVs at low-altitude as a flying BS to meet traffic demands temporarily of a hot-spots area. One of the popular commercial application areas of UAV with cache-aided NOMA technology is depicted in Fig. \ref{Fig.Application_UAV}, where a large number of mobile users in a hot-spot area are under the coverage of a ground macro base station (MBS). The MSB is overloaded and unable to satisfy the users’ requirements during peak hours because of the limited available frequency band. Cache-enable UAVs are deployed to assist the MBS in delivering users’ requested files. The battery-operated UAVs and the MBS are connected through wireless channels. When the battery is exhausted, it is recharged, or the UAV is replaced by a new one.

\begin{figure} [t]
	\begin{center}
		\includegraphics[scale=0.70]{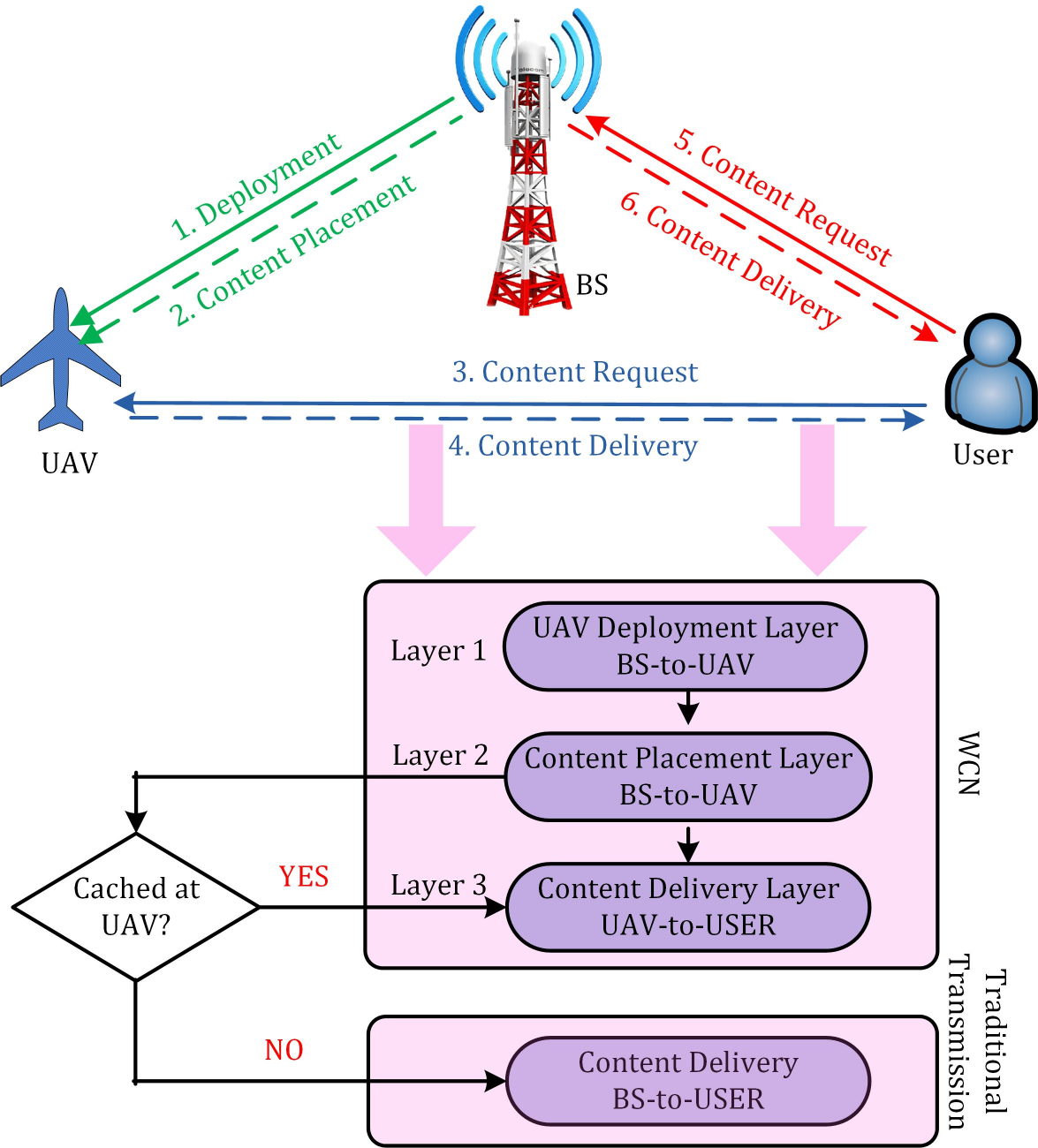}
		\caption{\small Cache content delivery strategy and different layers of a UAV-assisted wireless caching system. User requests for the desired file to the BS only when the file is not cached \cite{yin2021cross}.  }
		\label{Fig.cache_UAV}
	\end{center}
	\end{figure}
The UAV enabled cache-aided NOMA network is divided into three layers, (i) UAV deployment layer, (ii) content placement layer, and (iii) content delivery layer \cite{yin2021cross}, as shown in Fig. \ref{Fig.cache_UAV}. In the content placement layer, the MBS downloads popular content using the backhaul link and stores it in the cache of UAV using NOMA. The caching contents are replaced/updated regularly. In the content delivery layer, UAV groups users to deliver the requested contents using NOMA based on the CSI. A statistic QoS-based fixed (SQF) and instantaneous QoS-based adaptive (IQA) power distribution methods are applied in a UAV-enabled cache-aided NOMA system to improve the outage probability performance. Furthermore, an improved power allocation strategy named cross-layer based optimal method is employed to maximize the system hit probability \cite{yin2021cross}. A deep reinforcement learning (DRL) algorithm is proposed for content placement and delivering in a cache-enabling  UAV-assisted cellular network \cite{ wang2020deep}. The cache-enabled UAV serves users directly on the availability of the requested file in the cache. Otherwise, user requests for the files to the MBS directly \cite{yin2021cross} or via UAV \cite{ Zhang2020Cache}.

The resource allocation in UAV with cache-aided NOMA system has been studied in \cite{wang2020caching, Zhang2020Cache,  Zhang2020Caching, Thanh2020UAV }. In \cite{ Zhang2020Cache}, resource allocation for a UAV-assisted cellular system has been considered for maximizing the quality of experience (QoE) of the users by optimizing the content placement in the cache, location of UAV, and user association. In \cite{ wang2020caching }, to minimize the delivery delay, the authors have modelled an optimization problem for UAV deployment, caching placement, and power allocation of NOMA as a Stackelberg game. However, to minimize the content delivery delay, the authors in \cite{Zhang2020Caching} have incorporated the Markov decision process for jointly optimising the content placement, user scheduling, and power allocation to NOMA users. Increasing the operation time of battery-operated UAVs is one of the challenging issues. The energy spent for cache content placement and replacement further reduces flying time. To prolong the operation time of UAVs, the authors in \cite{Thanh2020UAV} have deployed a UAV that can harvest solar energy from the environment. Various algorithms researches have been applied to solve the optimization problem of UAV systems which have hardly considered the dynamic networks environment including the movement of UAV. The authors have applied a Markov decision process (MDP) to model caching placement and resource allocation with dynamic UAV locations and content requests \cite{ wang2020caching } \cite{Zhang2020Caching}.

In  \cite{Zhang2020Cache} and \cite{ Zhang2020Caching}, authors have proposed a UAV-assisted framework for delivering multimedia contents to the users located in a hotspot area. Here, cache-aided mobile UAV operated as a BS that reduces the backhaul link traffic providing the cached contents to the users’ group by NOMA. In \cite{dai2020uav}, Haibo \textit{et al.} have developed a cache-aided UAV-assisted vehicle-to-network (V2N) communication system where the UAV operates as a flying base station to communicate with vehicles. Cache-aided UAV was deployed to maximize the sum fairness of the vehicles. In \cite{wang2020deep}, cache-enabled UAV was deployed in a cellular network to assist the delivery of the user requested multimedia contents. Cache-enabled UAV was deployed in a NOMA-based MEC network to minimize the consumption of total energy in \cite{ budhiraja2021energy}. 

\begin{figure} [t b!]
	\begin{center}
		\includegraphics[scale=0.60]{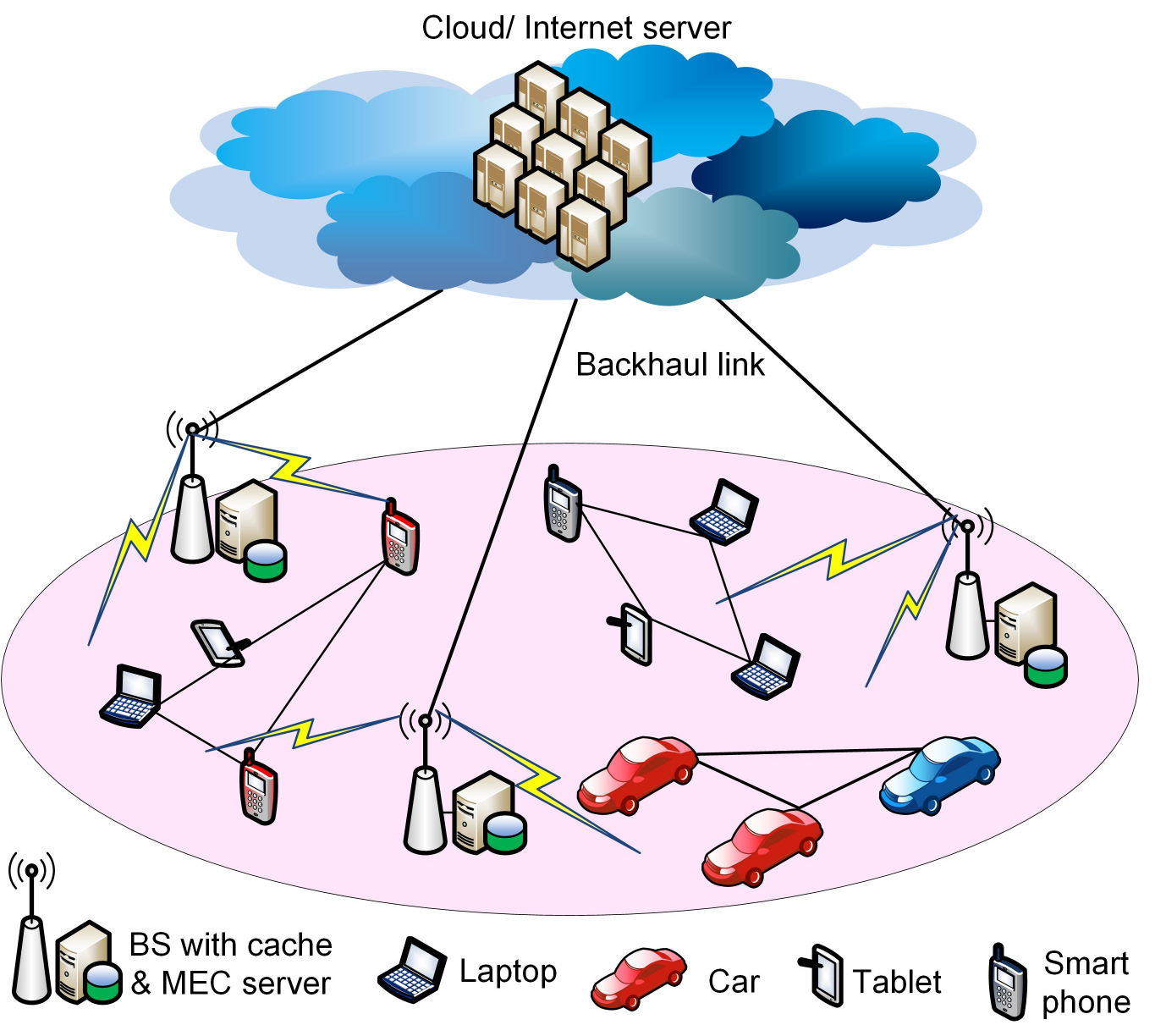}
		\caption{\small  Application scenario of cache-aided NOMA MEC networks. Users are connected and cooperatively sharing files. Devices offload their computational task to the MEC-cache-aided BS.}
		\label{Fig.BasicMEC}
	\end{center}
	\end{figure}

\subsection{Mobile Edge Computing}
Mobile edge computing (MEC) improves the cloud computing capability by shifting computing facilities at the edge of highly latency-sensitive networks such as cloud gaming and multiplayer gaming, autonomous vehicle functions, real-time drone detection, etc. In addition, caching in the MEC server further enhances the quality of communications and reduces backhaul load. Offloading the computation workloads of the mobile users, MEC assists the existing applications to improve their performance in terms of congestion in networks, delivery latency, and QoE. Cache facilitated MEC significantly intensifies the performance further \cite{Hao2018Energy}. Generally, the nearest APs to the users are equipped with cache-enabled MEC servers. The users within the coverage area of an AP get access to the caching contents that significantly reduce the backhaul link traffic and data transmission rate. The NOMA strategy empowers MEC to cope with the massive connectivity and huge data traffic of mobile users. NOMA-based MEC (NOMA-MEC) networks are capable of offering flexible computing services to mobile users. Some of the applications of cache-aided NOMA-MEC networks are depicted in Fig.\ref{Fig.BasicMEC}. The task caching in the MEC refers to storing some of the popular completed tasks and their associated data in the cache. Unlike the other cache-aided applications, in MEC, task caching requires computation in addition to storage. The user requested task can be computed locally in the mobile device or offloaded to the MEC server.

\begin{figure} [t b!]
	\begin{center}
		\includegraphics[scale=0.50]{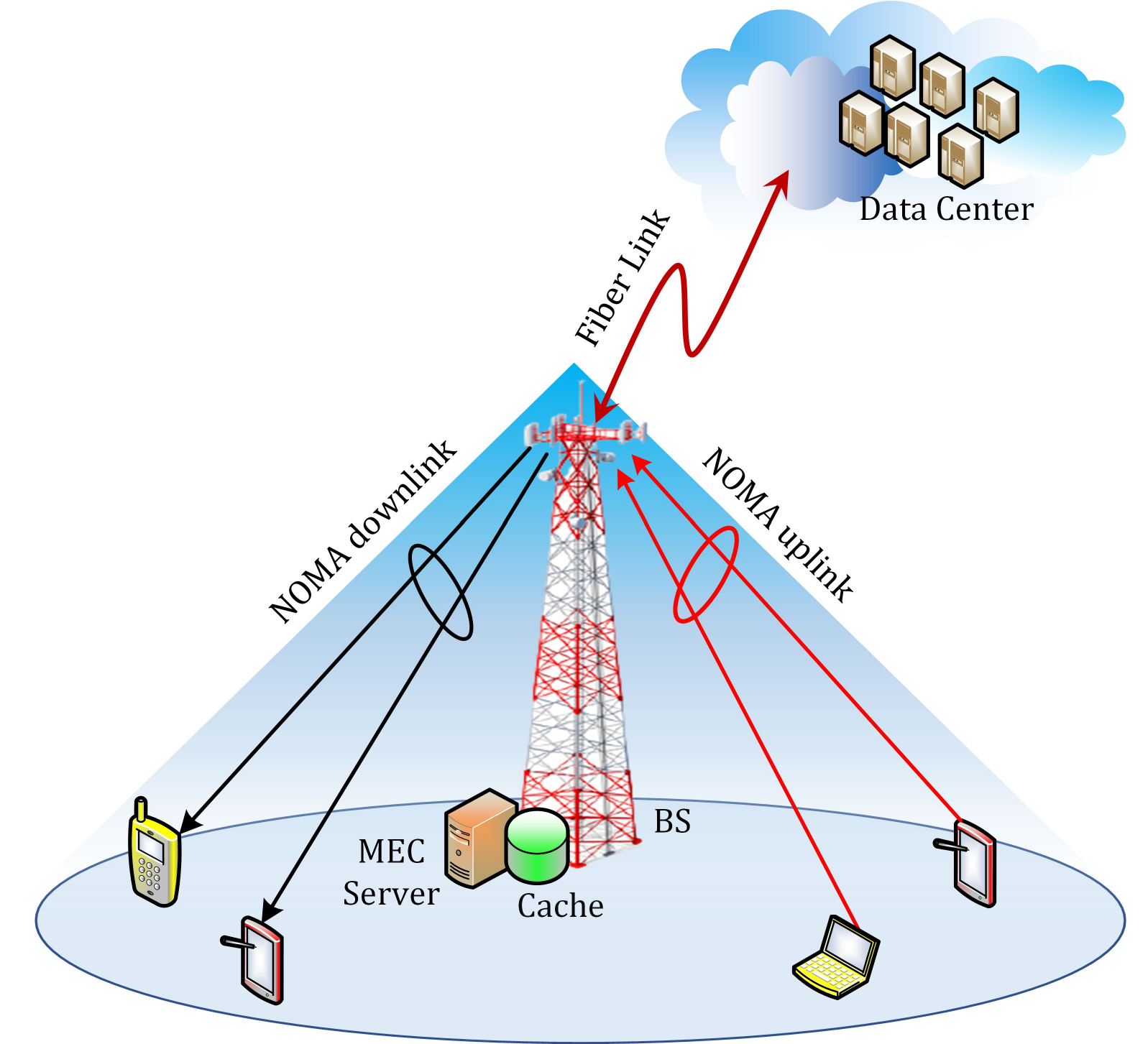}
		\caption{\small  Basic architecture of a cache-aided NOMA MEC network. User's devices offload computational tasks to cache and MEC-aided BS \cite{huynh2021joint}.}
		\label{Fig.cacheNOMA_offload}
	\end{center}
	\end{figure}

\textit{Local Computation:-} Consider a simple edge caching NOMA-MEC network with a BS, $N$ mobile users and a remote cloud, as shown in Fig. \ref {Fig.cacheNOMA_offload}. The BS is equipped with a MEC server with a finite storage capability. Let $L_n$ is the size of the requested content, $f_n^l$ is the local computing capability, $W_n$ is the number of cycles required to finish the given task, $R^{BH}_n$ and $R^{DL}_n$ are the average data transmission rate through backhaul link and downlink NOMA, respectively. $\mathcal{T}^{Lo}_n$, the total computational latency for executing a task locally includes  backhaul latency ($\mathcal{T}^{BL}_n= L_n/ R^{BH}_n$), downlink latency ($\mathcal{T}^{DL}=L_n/R^{DL}_n$) and local processing time ($\mathcal{T}^{LP}_n= W_n/f_n^l $) which is given by 
\begin{equation}
    \mathcal{T}^{Lo}_n=  (1-C_n)\mathcal{T}^{BH}_n + \mathcal{T}^{DL}_n +\mathcal{T}^{LP}_n,
\end{equation}
where $C_n$ is the status of the cache content, $C_n=1$ if the requested content is available in the BS cache, else it is $0$.
\textit{Edge Offloading:-}The primary objective of the MEC is to offload the computational task to the BS as much as possible for the remote execution. The total computational latency for edge offloading includes the uplink transmission time, processing time of the MEC server, and the backhaul delay. If a $U_n$ is the data of the offloaded task to the BS by the \textit{n}th user and the $R_n^{UL}$ is the average NOMA uplink data rate, the uplink delay is given as $\mathcal{T}^{UL}_n=U_n/R_n^{UL}$. The edge processing time for the offloaded task is $\mathcal{T}^{EP}_n= W_n/f_n$, where $f_n$ is the resources allocated by the MEC server for executing the computational task. Now, the total edge offloading latency is expressed as
\begin{equation}
    \mathcal{T}^{EO}_n=  (1-C_n)\mathcal{T}^{BH}_n + \mathcal{T}^{UL}_n +\mathcal{T}^{EP}_n.
\end{equation}

Since the size of the result of the offloaded task is much smaller than that of the offloaded task, and the downlink data rate is much higher than that of the uplink (as the transmitting power of the BS is higher than that of the users' device), the downlink time delay and the energy consumption associated with the downloading the result of the offloaded task can be neglected \cite{wang2017joint}. Hao \textit{et al.} studied the challenges related to the joint optimization of the task caching and the offloading in \cite{ Hao2018Energy }. Based on the popularity of the historical tasks, the LSMT algorithm predicts the future task popularity as a function of time. When the popularity of the computational tasks is unknown, the Gated Recurrent Unit (GRU) algorithm can be applied to predict it for a time-varying system \cite{ li2020joint}. Depending on the predicted popularity of the task, a multi-agent Deep-Q-network (MADQN) algorithm was applied to deal with the problem of caching and offloading. A new collaborative task offloading scheme proposed in \cite{ Yu2018Computation } is capable of reducing task execution delay up to 42.83\% a for single-user caching-enhancement scheme. 

Various types of resource allocation methods for cache-aided MEC with NOMA are found in the literature \cite{Yang2020Cache, Yang2020Distributed, budhiraja2021energy, rezvani2019cooperative, Qian2020NOMA }. For efficiently completing the computation tasks of the users, a resource allocation optimization problem under the constraints of caching and computing resources is formulated and addressed by an SAQ-learning-based algorithm in \cite{Yang2020Cache}. In \cite{ Yang2020Distributed } also, authors have applied the SAQ-learning-based method to solve the problems associated with the optimization problem for minimizing total energy consumption subjected to offloading decision, computation resource, and caching decision. In \cite{budhiraja2021energy}, the authors designed a MEC network where UAV is deployed as a moving edge cloud server to offload the computation workloads of the mobile terminals. A resource allocation framework for video caching placement and delivery was developed in heterogeneous cache-aided MC-NOMA networks \cite{rezvani2019cooperative}. The delivery-aware cache placement strategy (DACPS) jointly allocates physical and radio resources during the cache placement phase, and the delivery-aware cache refreshment strategy (DACRS) deals with the dynamic behavior of the channel during the delivery phase \cite{rezvani2019cooperative}. Incorporating the overall tasks completion delay and total consumption of computational resources by the edge servers, the authors formulate a system cost function. Thereafter, jointly optimize the computational resource allocations at edge servers and radio resources for smart terminals to minimizes the system cost function \cite{ Qian2020NOMA }. The authors in \cite{Zhang2021Distributed} formulate a new utility function considering offloading time, available resources, and caching decision, and maximize it subjected to the transmission bandwidth, available computing resources, and storage resources. In \cite{ huynh2021joint}, the authors aimed to reduce the total completion latency for all users of a cache-aided NOMA-MEC, formulated a joint optimization problem of offloading decision, caching strategy, computational resource, and power allocation under the constraint of energy consumption, offloading decision, and computation and storage capacity. In \cite{Wu2018NOMA} also, the computation delay to finish mobile users’ tasks was minimized by jointly optimizing the offloaded workloads and data transmission time.

\subsection{V2X Communication}
Vehicular communications have gained a huge attraction among researchers due to the possibility of improving travel experience in terms of road safety, internet access for on-board information, and entertainment facilities. The IEEE 802.11p technology-based communication for vehicular ad hoc networks (VANET) provides 6 - 27 Mbps data rate for a short distance communication \cite{ Jiang2008IEEE}. LTE-based vehicle-to-vehicle (V2V) communication supported by the Third-Generation Partnership Project (3GPP) also emerges as an efficient approach \cite{Chen2016LTE}. The V2V communications not only provide an entertainment facility to the onboard user but also provide safety, traffic information, pollution control, and traffic applications that require a huge amount of data transmission. In addition, the short duration of connectivity between vehicle-and-infrastructure (V2I), frequent change of channel gain quality, and fast movement of the vehicle make V2V communication further challenging. Liang \textit{et al.} have studied the fundamental challenges to empower efficient vehicular communications from the physical layer perspective in \cite{ Liang2017A }. In \cite{ fang2020content }, the authors verify the superiority of NOMA over conventional OMA in terms of enhancing the content delivery efficiency. Two dynamic cache content placement schemes are proposed for adaptive bitrate streaming of video in vehicular communications \cite{ Guo2018Cache }. The NOMA technique is well recognized in vehicular communication for the capability to handle massive connectivity and outperforms the traditional OMA-based system \cite{ Di2017NOMA }. The authors in \cite{ Di2017V2X } investigated the spectral efficiency and resource allocation of a NOMA-based vehicular system.

In V2V communication, cache-enabled vehicles store some of the popular contents. Vehicles communicate with the BS during the cache placement phase and on the unavailability of the requested file in the cache of neighboring vehicles. To understand the working principles of cache-aided NOMA in V2V communication, consider a simple vehicular communication model consisting of two cache-aided vehicles ($V_1$ and $V_2$) and a BS. Let users of $V_1$ and $V_2$ request for files $f_1$ and $f_2$ respectively. Consider an extreme case when neither $f_1$ nor $f_2$ is cached. The BS downloads the files from the internet using the backhaul link and then transmits files applying NOMA. The traffic load of the backhaul link is the same as of the conventional NOMA. When any one of the files (say, $f_1$) is cached, one user ($V_1$) receives its file from a neighbor ($V_2$), and another user ($V_2$) gets its file from BS (using backhaul link). Interestingly, for this case, interference can be removed completely, and users do not need to use SIC to decode their signal as the conventional OMA technique emplied to transmit both signals utilizing the whole available bandwidth. Hence, the average performance will be better than that of the conventional NOMA. Another extreme case is when both files $f_1$ and $f_2$ are cached. Here interference can also be avoided completely, and BS does not need to use the backhaul link. Hence, the average performance of the cache-aided NOMA in V2V communication will be significantly better than that of the conventional NOMA.

Gurugopinath \textit{et al.} in \cite{gurugopinath2019cache} first proposed a cache-aided NOMA in vehicular communication. The authors consider full file caching and split file caching techniques in vehicular networks. In full file caching, each vehicle stores and requests entire files following NOMA principle; whereas the split file caching technique divides content into two parts. The challenges of cache-aided NOMA in vehicular communications addressed in \cite{ gurugopinath2020non }. A hybrid multicast/unicast scheme has been investigated in cache-aided NOMA-based vehicular networks \cite{ pei2020hybrid }. In order to maximize the unicast sum-rate, the authors have formulated an optimization problem subjected to peak the transmit power, backhaul capacity, the minimum unicast rate, and the maximum multicast outage probability. Chao \textit{et al.} in  \cite{ li2020cache } have studied the cache-assisted physical layer security of NOMA-based vehicular communications.

\subsection{Cell-free Massive MIMO}
 Unlike conventional cellular topologies that mainly serve human users, next-generation communication systems provide mMTC also. The cellular networks cannot handle connectivity to billions of user terminals. Therefore, a cell-free communication topology with decentralized technology is required for next-generation communication, and cell-free massive MIMO (CFmMIMO) technology could be a potential approach \cite{Zhang2019Cell}. The CFmMIMO comprises large numbers of low costs and low operating powered AP antennas distributed over a wide area and coherently serves user terminals by all nearby AP antennas. A front-haul network connects all the antennas of the APs to central processing units connected with internet servers through a backhaul network. Through this network design, user terminals get AP antenna very close to them, and consequently, achieve improved QoS. Though both the CFmMIMO and distributed massive MIMO can serve thousands of user terminals, they are different. In distributed massive MIMO, BS antennas are installed over a cell and serve only the users within that cell. On the other hand, CFmMIMO is free from a geographical boundary, and antennas serve all users. 
 
 Primarily, NOMA was employed in CFmMIMO to reuse pilot sequences within the same cluster which significantly serves more users than the conventional OMA \cite{ Zhang2019Spectral}. However, under the low number of active users scenario in a CFmMIMO system, OMA is superior to the NOMA in achieving sum-rate because the NOMA suffers from intra-cluster pilot contamination and imperfect SIC \cite{ Li2018Cell }. To deal with this an adaptive NOMA/OMA mode-switching method is proposed in \cite{Bashar2020Performance, Bashar2019NOMA}. The phase-related mismatch at the AP degrades the spectral efficiency of a NOMA-based CFmMIMO system. However, NOMA is still capable of outperforming OMA under mismatches and imperfect SIC \cite{Ohashi2021Cell }. The authors have applied Poisson point processes in the NOMA-based cell-free massive MIMO to model the random user and AP locations and found NOMA to be an efficient technique in enhancing the overall rate, especially under low path loss exponents and high AP densities \cite{ Kusaladharma2019Achievable, Kusaladharma2021Achievable }. 
 
 CFmMIMO is viewed as one of the new technologies for 5G and B5G communications due to its uniform service quality, robust diversity, and interference management ability \cite{Zhang2019Cell, Zhang2020Prospective}. Distributed APs of CFmMIMO make it suitable for caching. Integrating with CFmMIMO, caching reduces backhaul traffic load and energy consumption significantly. Recently Chen \textit{et al.} \cite{Chen2021Wireless} have introduced a cache-aided CFmMIMO framework in 2021. However, in the year 2018, the authors have applied a coded caching in a cell-free environment and derived analytical expressions ergodic spectral efficiency and outage probability expressions for analyzing the performance of SIMO network \cite{Mozhgan2018Coded}. Wang \textit{et al.} proposed a smart caching scheme in MEC-enhanced small-cell Massive MIMO networks, where MBS and SBSs are equipped with caching memory \cite {Wang2020Framework}. Based on the user’s request history, the MEC server can predict the next content that might be requested and starts caching that content in the SBS. Chen \textit{et al.} \cite{Chen2021Wireless} compared the performance of a CFmMIMO with small cells from caching strategies perspective and established CFmMIMO as a superior technology over small cells in terms of the successful content delivery probability and total energy consumption.
 
\subsection{Wireless Powered Communication}
Simultaneous wireless information and power transfer (SWIPT) through dedicated radio frequency emerged as a superior wireless energy harvesting (EH) technique that prolongs the battery life and provides uninterrupted network operation. Maximizing the energy efficiency is one of the fundamental objectives of 5G networks. Using the time switching (TS) or power splitting (PS) protocol, the SWIPT-enabled system extracts both information and energy from the ambient radio signals simultaneously. Consequently, SWIPT improves the energy efficiency of wireless systems \cite{Feng2020UAV}. The combined NOMA-and-SWIPT-based paradigms enhance spectral efficiency and energy efficiency of 5G systems, and support the services of the IoT and the mMTC \cite{Tang2019Energy}. Wu \textit{et al.} designed a transceiver for NOMA-based SWIPT-enabled cooperative full-duplex relaying systems \cite{ Wu2019Transceiver}. Yuan \textit{et al.} considered a cooperative NOMA transmission scheme in a PS-based SWIPT system and formulated an optimization problem that maximizes energy efficiency and reduces the energy consumption of the system, especially in the low power region \cite{Yuan2019Energy}. A few articles also found in the literature where the NOMA has been adopted in SWIP system and proposed various methods to analyze different metrics such as outage probability, throughput and energy efficiency \cite{ Li2020Performance, Do2018Optimal, liu2016cooperative }.

Caching is popularly applied in sensor networks to reduce energy consumption and improve the energy efficiency of sensors. An AP employed as a gateway to sensors is facilitated with cache memory stores the sensing data temporarily and updates it periodically. The gateway retrieves cached sensing information and delivers it to multiple users without activating the sensors frequently (which consumes substantial energy). In \cite{Niyato2016novel}, the authors have introduced caching in IoT sensing services and proposed a caching mechanism in EH-enabled sensor networks that improves the sensing performance significantly. The impact of caching and EH on the energy consumption at small cell base stations has been investigated in \cite{kumar2015tradeoff } and shown that instead of existing trade-off between the size of cache and harvesting equipment at the SBS, caching achieves desired system performance. In \cite{ Zhou2015GreenDelivery}, the authors have proposed a new network paradigm named as GreenDelivery, where based on the popularity and harvested energy EH-enabled small cells cache and push the multimedia contents before it is requested. This network framework considerably reduces macro-BS activities and consequently decreases energy consumption.

In addition to the above articles, researchers have integrated the NOMA technique in cache-enabled EH networks. The caching in the NOMA-enabled SWIPT model functions efficiently in the enhancement of the quality of user experience (QoE) \cite{li2021energy}. The authors have proposed a joint content push and transmission scheme in a cache-enabled SWIPT-based relying network \cite{zhang2020joint}. With the help of the NOMA technique, a two-stage content push and the delivery scheme has been proposed to achieve superior spectral efficiency. Another joint content caching and EH method is proposed to improve the performance of EH and information transmission for a NOMA-based IoT network in \cite{li2021energy}. Cache-aided NOMA is not explored much yet in the EH-enabled networks. However, researchers are implementing cache and NOMA separately and cache-aided NOMA primarily to enhance energy efficiency, QoE and reduce the energy consumption of SWIPT networks. 

\subsection{mmWave Communication}
Millimeter Wave (mmWave) band ranges roughly from 30GHz to 300GHz, providing an alluring spectrum bandwidth of 270GHz. Compared to existing wireless technologies, mmWave proves advantageous in terms of available bandwidth, size of elements, and narrowed beams \cite{WangmmWave}. Despite all the privileges that mmWave technology offers, it is very challenging for the practical implementation mmWaves for 5G networks. The prime reason behind the shortcomings of mmWaves is it is channel characteristics tend to have high path loss, significant atmospheric absorption, and have difficulty in non-line-of-sight communication \cite{WangmmWave,Rappaport5Gcell}.

While dealing with mmWave communications, one of the drawbacks is multiple access because of high power consumption and costly hardware \cite{ZhummWaveNOMA}. MmWave, when integrated with NOMA, can overcome this limitation as NOMA can provide access to multiple users simultaneously in Power Domain. Not only can it increase the number of users, but also it can contribute to better data rates and reduced interference \cite{NaqvimmWaveNOMA}. In \cite{ZhangmmWaveMMIMO}, when compared with existing LTE systems, a significant capacity improvement was achieved on combining mmWave-NOMA with massive MIMO systems. Comparative analysis of the Outage probability for downlink NOMA-mmWave network over small-scale fading channels concluded that NOMA outperforms OMA in multi-cell-based mmWaves systems\cite{SunmmWaveNOMA}. \cite{LimmWaveNOMA} suggested application of agile beam NOMA for mmWave networks and observes that NOMA-mmWave has better coverage probability and sum-rate than OMA mmWave networks. Compared with TDMA based MIMO-mmWave networks for D2D communications, for NOMA-based MIMO-mmWave networks, Outage Probability tends to decrease exponentially, whereas ergodic capacity increases linearly\cite{LiD2D}. All the before-mentioned literature used the Nakagami-m Fading model to represent small-scale fading, whereas \cite{TianmmWaveNOMA} used Fluctuating Two-Ray model (FTR) obtained the same results with a better precision. Thus, we can summarize that NOMA-mmWave based 5G networks can accomplish a better overall throughput than conventional OMA-based Networks.

\subsubsection{NOMA in mmWave}
Next, we will discuss some of the areas where integration of NOMA and mmWave can enhance the system throughput.
\begin{itemize}

    \item \textbf {Beamforming mmWave-NOMA}: For mmWave-NOMA, beamforming is achievable in two ways: single beamforming and multi-beamforming. In the case of multiple users, single beamforming is rather disadvantageous because it will require wider beams, reducing the beam gain. Thus, authors in \cite{ZhummWaveNOMA} suggested multi-beamforming for multi-users where the base station can simultaneously have multiple narrow beams directed towards multiple users. It will provide higher beam gain, robustness, and a better sum-rate. Despite its supremacy, mmWave-Noma with multi-beamforming has challenging Antenna Wave Vector (AWV) design due to Constant Modulus (CM) constraint. It occurs due to the presence of phase-shifters which are generally non-convex and high dimensional. Currently, research is going on in this direction, and further studies in this direction are required.

    \item \textbf {MmWave Massive MIMO}: With mmWave communications, using a large-scale antenna array is feasible, compensating for poor propagation conditions for mmWaves. The use of mmWave technology in massive MIMO enables low-cost-low-power components. However, the combination of mmWave and massive MIMO has its challenges. Pilot contamination, prior knowledge of accurate CSI, accurate channel estimation, channel feedback, and real-time realization are significant concerns \cite{BusarimmWaveMMIMO}. Since the mmWave massive MIMO channels are not independent and identically distributed (i.i.d), the channel model can no longer be realized as orthogonal practically. Thus, exploration of non-orthogonal techniques such as NOMA was encouraged\cite{BusarimmWaveMMIMO}. For a large number of users scenario, NOMA can further reduce the problem of pilot contamination using power domain NOMA and SIC\cite{LiNOMAMMIMO}. NOMA also reduces the outage probability and improves spectral efficiency and energy efficiency of the mmWave massive MIMO systems \cite{GhoshNOMA}. MmWave-NOMA system needs the channel information at the base station, which produces significant overhead data. We can expect this overhead to increase further with the inclusion of MIMO, and with massive MIMO, this overhead data would be very large\cite{AghdammmWave}. Thus, a detailed study on this aspect of the mmWave-massive MIMO-NOMA system is needed.

    \item \textbf {Cognitive mmWave Networks}: Cognitive Radios (CR) are the Dynamic Spectrum Access Networks where a licensed primary user shares spectrum with the unlicensed secondary users for transmission. Underlay CR networks enable the secondary user to transmit the data in the presence of the primary user on the condition that secondary user transmission power is low compared to that of the primary user. The Secondary user thus cannot transmit over a long range. Since mmWaves also work on short-range, mmWaves can be the enabling technology for CR systems to provide higher capacity and data rate at increased spectrum efficiency \cite{SongCRmmWave,SahaCRmmWave}. Since power-domain NOMA can easily do justice on power handling for both primary and secondary users, integration of NOMA with Cognitive mmWave Networks has good potential. \cite{SongmmWaveNOMACR,Songpower} has discussed the security aspects of NOMA-based Cognitive mmWave Networks, but other aspects are still unexplored.

\end{itemize}

\begin{figure} [t]
	\begin{center}
		\includegraphics[scale=0.45]{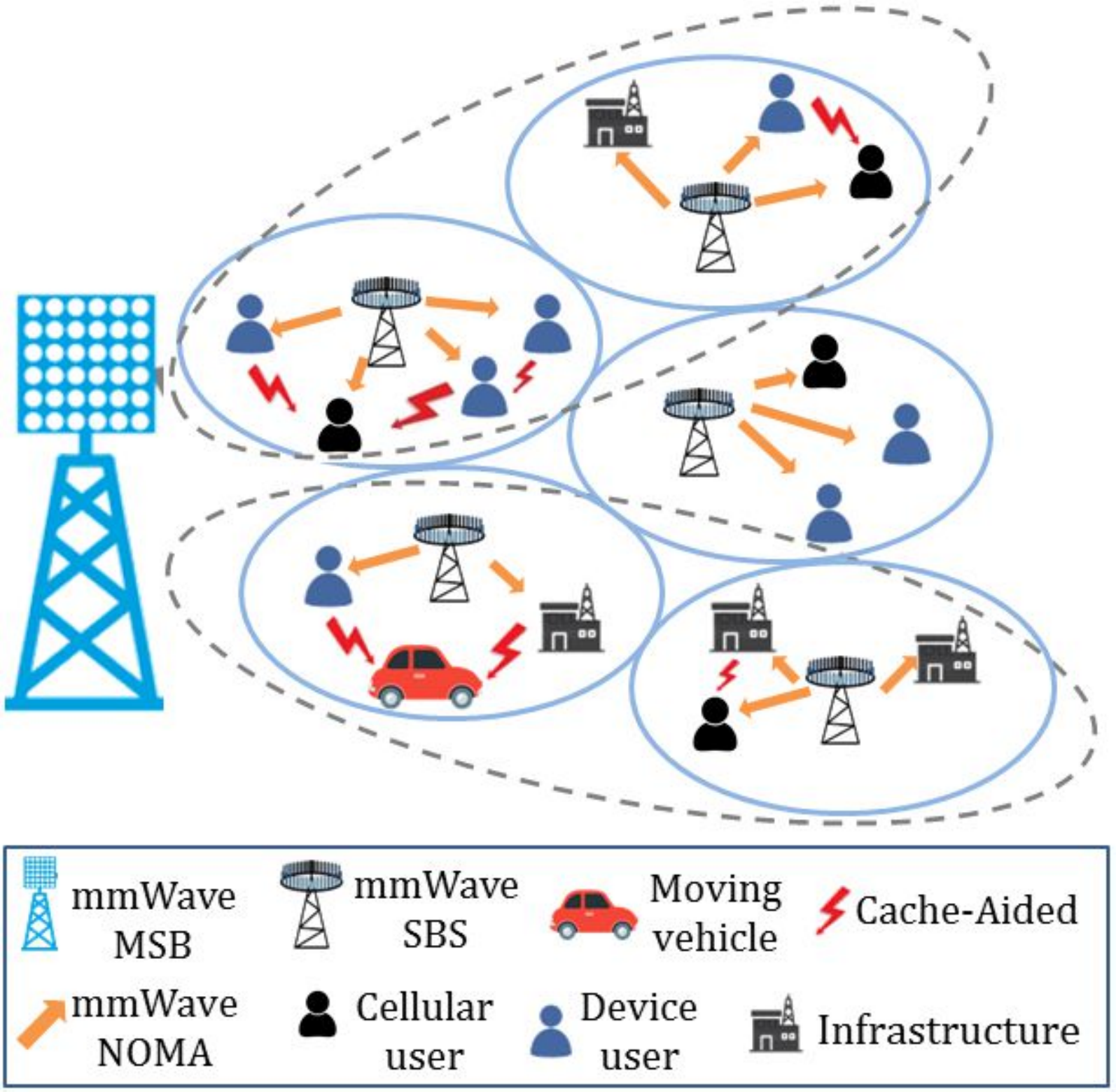}
		\caption{\small Cache-Aided NOMA for mmWave.}
		\label{Fig.Cache-Aided NOMA for mmWave}
	\end{center}
	\end{figure}
\subsubsection{Caching with mmWave NOMA}
MmWave-NOMA requires a considerable amount of backhaul overhead. Cache-aided systems can overcome this problem by saving the contents on the cache-enabled user device and small base stations. Caching for millimeter waves was suggested in \cite{SemiariCachemmWave}, where the authors acknowledged that the use of cache in mmWaves could reduce frequent handovers and handover failures. Authors exploited the high storage capacity of the modern smartphone to store the data in mobile user equipment (MUE) and retrieve using high capacity mmWaves when required. It eases the overhead backhaul, especially in fast-moving MUEs, alleviates service delay problems.

Although the research community appreciated the use of cache in mmWave and NOMA; cache-aided mmWave NOMA is not adequately addressed. In this article, we propose a mmWave NOMA system that uses caching to reduce the backhaul data. MmWaves at 28GHz can have a larger coverage area than 60GHz as in prior frequency mmWaves get less attenuated. In Fig. \ref{Fig.Cache-Aided NOMA for mmWave} we have considered a mmWave macro base station (MBS) antenna transmitting at 28GHz using NOMA aided beamsteering. Each cell contains a small base station (SBS) antenna to receive the signals from the macro base station. Users of each cell get connected to SBS for communication. These users can be pedestrians, any infrastructure, vehicle, device, grid, etc. Also, these users may or may not be cache enabled. Due to the large storage capacity in modern-day devices, these devices can act as a cache-enabling platform. When a non-cached user requires specific data like a music file, the user sends its request to its SBS. SBS, in turn, asks the cache-enabled devices where the data gets stored in cache memory based on the principle of popularity. If the required file is available in any cache-enabled devices, the device sends the file to SBS, which forwards the file to the requesting user. The process will reduce the time and backhaul networking required by the SBS to reach out to MBS, which downloads the music file from the server. If the data is not available in the cell, the SBS may connect to nearby SBSs for the content, successively asking cache-enabled devices in their coverage area. If available, the content gets transferred through backhaul networking. If the content is not available in any cache-enabled devices, the SBS will request the content to MBS.
\begin{figure} [t b!]
	\begin{center}
		\includegraphics[scale=0.47]{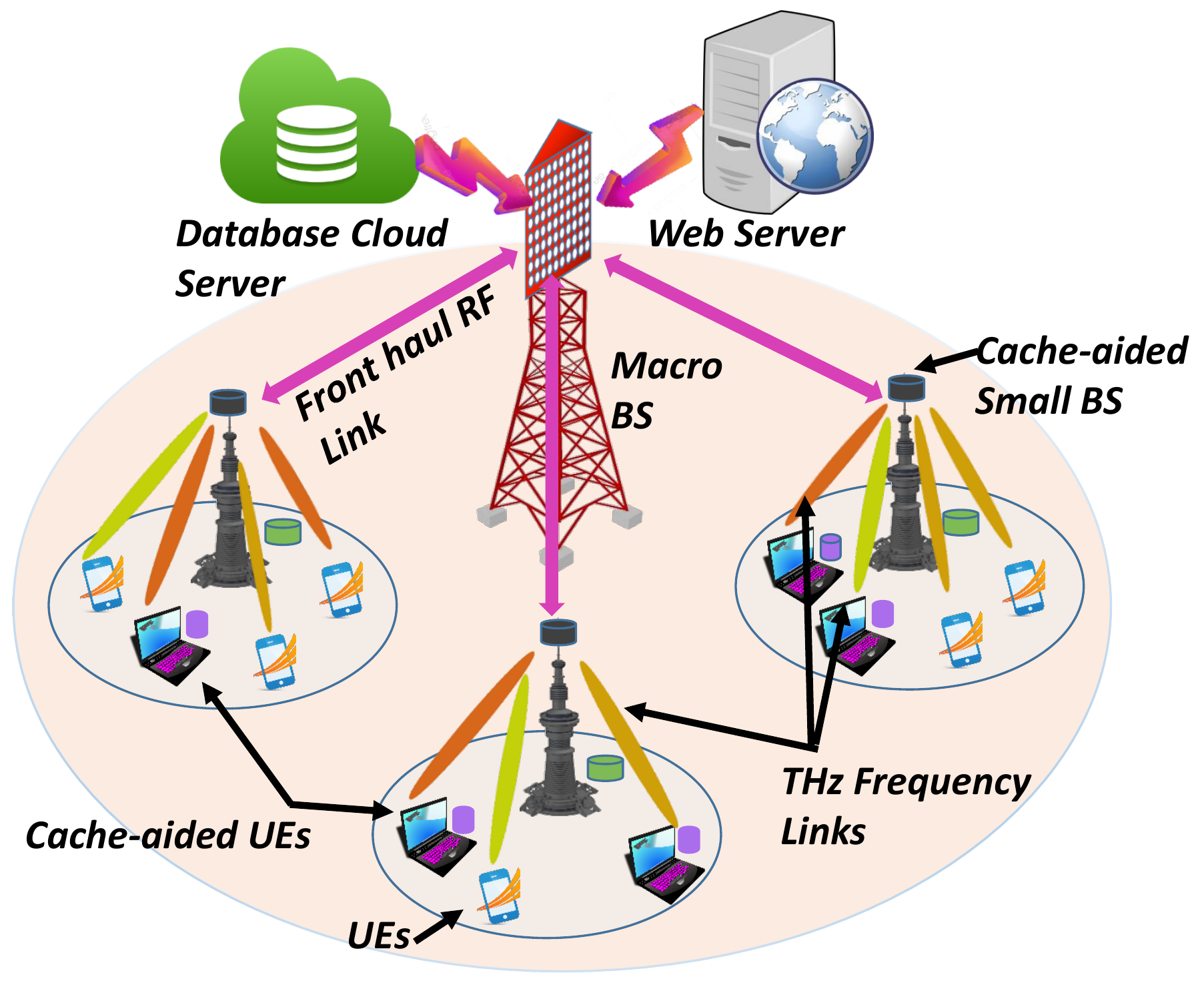}
		\caption{\small Scheme for THz based Cache-Aided NOMA.}
		\label{Fig.Cache-aided THz-NOMA}
	\end{center}
	\end{figure}
\subsubsection{Caching with Tera Hertz-NOMA}
Compared to mmWave bands, THz bands provide higher bandwidth, lower eavesdropping and less free-space diffraction\cite{ElayanTHz}. Even at a distance of 5 meters, the NLOS THz propagation channel capacity reaches 100Gbps\cite{MoldovanTHz}. However, high frequency selective path loss, non-existent multipath gains and a complex mMIMO system threaten THz deployment\cite{UlgenNOMATHz}. Studies suggest that introducing NOMA systems for THz bands can fill these gaps efficiently\cite{MelhemNOMATHz,MagboolNOMATHz}. NOMA can reduce the processing complexity at the receiver, diversify highly correlated channels, and increase user fairness\cite{MagboolNOMATHz}. Ulgen \textit{et al.} \cite{UlgenNOMATHz} showed that for a 350 GHz downlink, NOMA outperforms OMA in terms of the data rate. Further, the NOMA-based THz system has a data rate three times higher than OMA for the same distance and environmental conditions. Although NOMA-THz systems outsmart OMA-THz systems, studies suggest that NOMA-THz systems lags behind NOMA based mmWave/microwave systems.\par
\begin{figure*} [ht!]
\begin{center}
\includegraphics[scale=0.4]{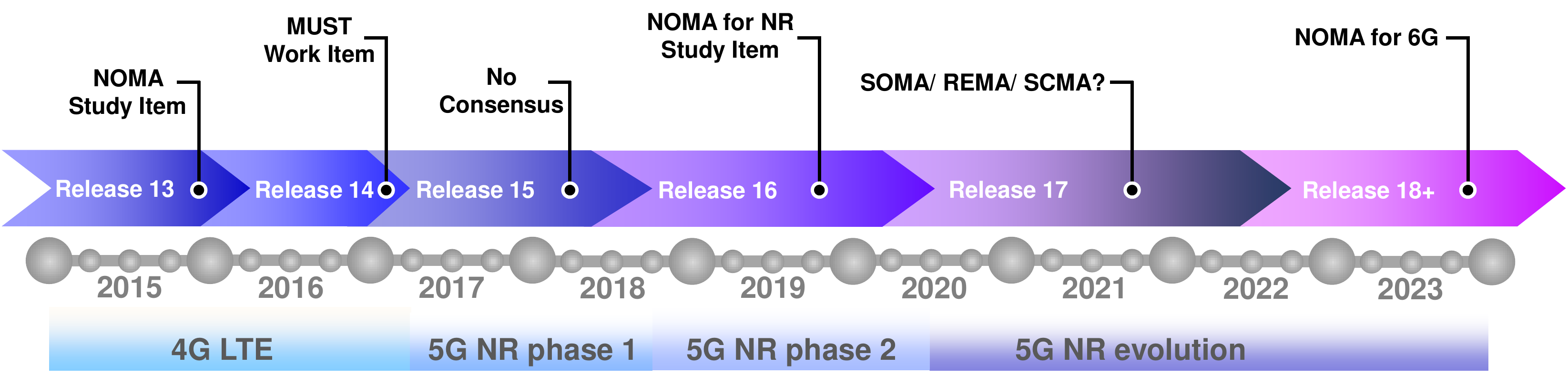}
\caption{\small Timeline of $3$GPP releases and different NOMA related standardization activities. NOMA was included in one of the work items in release 16. In release 13 and release 16, discussions were limited to study items only.}
\label{Fig.TimelineStandard}
\end{center}
\end{figure*}
In \cite{SABUJTHzNOMA}, authors showed that for a BS-mMTC communication, the spectral and energy efficiency for THz falls below microwave and mmWave. For the THz-NOMA, a high number of users (more than two) in a beam cluster can cause significant user interference causing degradation in received signal quality and delayed SIC. Also, optimizing the power allocation can become non-convex and computationally complex \cite{MagboolNOMATHz}. To mitigate the problem of spectral efficiency, Zhang\textit{ et al.} \cite{ZhangTHzMIMONOMA} recommended a hybrid MIMO-NOMA two-tier architecture where the MBS-SBS link is RF-based, and SBS-UEs links are THz based. Since RF-based networks can cover a larger area, multiple cache-aided SBS are connected to a single MBS. THz signals cannot travel long distances; thus, the SBS will have smaller coverage but a very high data rate compared to RF counterparts. However, such an arrangement will result in a data rate mismatch between the two tiers. This mismatch can be reduced by utilizing the available cache at the SBS and in the UEs. Fig. \ref{Fig.Cache-aided THz-NOMA} shows that one can combine the cache memory and NOMA to generate THz-based cache-aided NOMA networks. As discussed in previous articles, cache-aided NOMA can increase the spectral and energy efficiency. There are still wide open research questions and a detailed study on THz-based cache-aided NOMA is lacking.

\subsection{Intelligent Reflecting Surfaces}
Intelligent Reflecting Surfaces (IRS) is an intelligent metasurface manufactured of a large number of programmable metamaterials. An IRS can mitigate the wave propagation blockage problem, enhance signal power, and suppress interference by dynamically adjusting the phase and polarization of the incident wave \cite{ Zheng2020Intelligent, gong2020toward}. IRS is envisioned as a potential technology for 6G because of its ability to improve the QoS of wireless networks by customizing the wireless propagation environment. The IRS technique constructively tunes the channel vector of users and boosts the advantages of implementing the NOMA transmission technique. In \cite{ Ding2020OnThe }, the authors have validated that an IRS-enabled NOMA outperforms the IRS-enabled OMA in terms of outage probability. The IRS technique mostly implemented in NOMA systems to further increase the coverage \cite{ Ding2020Simple}, energy efficiency \cite{fang2020energy }, sum-rate \cite{ zeng2020sum, Mu2020Exploiting }. Ding \textit{et al.} proposed an IRS-assisted NOMA transmission such that the network can serve more users compared with spatial division multiple access \cite{ Ding2020Simple }. Article \cite{chen2021exploiting} introduces an IRS-aided edge caching system to realize the maximum benefit of caching. Here, the authors formulated a network cost minimization problem regarding backhaul capacity and the transmission power to optimize the content placement. Although no article has analyzed the performance of cache-aided IRS-enabled NOMA systems, we can further improve the performance of cache-aided NOMA networks by implementing IRS in the AP-to-user link.    

\section{The Road Ahead}\label{sec: open}
Cache-aided NOMA networks are constantly evolving and are getting diversified with advancement in other domains. In this section, we highlight the recent related standardization activities first, and then point out a few open research challenges.

\subsection{Standardization Activities}
A communication standard ensures interoperability among vendors, establishes conformity with local/ international regulations, boosts confidence of startups, and attracts venture capitals/ tech giants. Large investments demand a fixed turnaround time, which in turn, reduces the technology rollout phase. Although cache-aided NOMA is mostly a concept so far, the standardization activities is a proof that the concept will soon become a reality. For example, the discussions in the extreme high throughput (EHT) study group of IEEE $802.11$ regarding semi-orthogonal multiple access (SOMA) \cite{Suh2018soma} inspired a NOMA prototype based on software defined radio (SDR) for Wi-Fi \cite{khorov2020prototyping}.  

Non-orthogonal multiple access has heavily influenced multiplexing schemes used for digital television (DTV). The advanced television systems committee (ATSC) $3.0$ standard uses layered division multiplexing (LDM) \cite{ATSC30phy}. LDM is a two-layer technique \cite{Zhang2016layered}. Unlike OMA, both of these layers use full frequency spectrum and full-time duration. Multiplexing between the layers is achieved through PD-NOMA, i.e., the layers are different at power levels. The upper layer has a higher power allocation and is meant for broadcast services to mobile terminals whereas, the lower layer has a lower power allocation and is used for multicasting to fixed reception terminals. While NOMA is fully incorporated in the American DTV standard ATSC, there had been many proposals to include NOMA in the other major DTV standard, Digital Video Broadcasting (DVB). These include the use of NOMA for satellite DVB-S2X \cite{Ramirez2020study} and for terrestrial DVB-T \cite{Vanichchanunt2021implementation}. A comprehensive account of NOMA based broadcast services may be found in the recent article by Shariatzadeh \textit{et al.} \cite{Shariatzadeh2021improving}.

Within the $3$rd Generation Partnership Project ($3$GPP), NOMA was first included in LTE Release 13 as a study item (SI) \cite{Benjebbour2015noma} and later, in Release $14$, NOMA was included as a work item (WI) \cite{ZTE2017noma} under the name of multi-user superposition transmission (MUST) \cite{3GPPtr36.859}. MUST is a grant-based downlink technique. Simulation studies showed that with MUST a $10\%$ to $30\%$ improvement in throughput is possible depending on other network deployment parameters. Release $15$ in mid-$2018$ marks the beginning of $5$G new radio (NR), which continued to advance in Release $16$ also known as $5$G NR Evolution. The grant-free uplink NOMA was included in Release $16$ SI \cite{3GPPtr38.812}. A complete summary of NOMA-based standardization activities within $3$GPP is available in the article by Chen \textit{et al.} \cite{Chen2018toward} and later by Yuan \textit{et al.} \cite{Yuan20205g}. On the other hand, the article by Cirik \textit{et al.} \cite{Cirik2019toward} discusses NOMA standardization in the larger grant-free landscape. Release $17$ is due on $3$rd quarter (Q$3$) of $2022$, and there had been some indications that NOMA, in its contention-based grant-free form, will continue to play an important role. The proposed variants include rate-adaptive constellation expansion multiple access (REMA) \cite{Perotti2015non}.

Despite the interest, NOMA was not adopted in any of the work items for $5$G NR, as seen in Fig. \ref{Fig.TimelineStandard}. Rather, OFDMA continues to be the downlink MA choice while single-carrier FDMA (SC-FDMA) has been finalized for uplink. This is because no consensus regarding NOMA could be formed by the large number of stakeholders \cite{Yuan2021noma}. However, as pointed out earlier, some variations of contention-based grant-free NOMA is being proposed for $6$G. In \cite{Yu2021sparse}, a variation of CD-NOMA, named as sparse-code multiple access (SCMA), is discussed. Exploiting the sparsity of the codebook matrix, the multi-user detection can be performed in SCMA with much lower complexity than maximum-likelihood detection.

NOMA has been the central topic for various companies' whitepapers \cite{Islam2017noma}. The first in the league is NTT DOCOMO, which published a series of technical documents \cite{DOCOMO2014noma,DOCOMO2018noma}. Huawei, in addition to NOMA and SOMA, proposed a third variant, rate-adaptive constellation expansion multiple access (REMA) \cite{Ding2016noma}. The interest in NOMA has been manifested by other large telecos (ZTE Corporation, SK Telecom) \cite{SKtel2014noma}, chip suppliers (Intel, Qualcomm) \cite{Qualcomm2018noma}, OEMs (LG Electronics, Samsung, Nokia) \cite{LGDing2016noma} and equipment vendors (Anritsu) \cite{Anritsu2016noma}.

Cache can aid NOMA based networks in multiple ways. One possible avenue is to reduce pilot overheads or completely get rid of pilots through prior statistical knowledge of data. Also, with prior knowledge it is possible to build connections keeping the radio resource control (RRC) in idle or in inactive state \cite{Ma2020lightweight}. There is a related two-step random access channel (RACH) standardization within $3$GPP as well \cite{3GPPrp190711}.

\subsection{Open Research Challenges}
\subsubsection{Joint Sensing and Communication Frameworks}
The 6G wireless communications systems are envisioned as joint radar sensing and communication paradigms, which simultaneously sense targets and communicate with the users. An integrated sensing and communication (ISAC) network empowered by cache-aided NOMA shares the spectral resources and infrastructure and could be an evolutionary framework for the next-generation communication systems. The communication signal can be exploited for target sensing by suitably designing the co-variance matrix of the transmitted signal \cite{Stoica20007Probing}. The cache empowered BS transmits a superimposed signal satisfying the necessary standard for target sensing hence, the superimposed signal can be exploited for communication and sensing also. The users employ the SIC process to recover their signals like the conventional process. The primary aim of the sensing system is to maximize the power of the probing signal towards the direction of targets \cite{Stoica20007Probing}. Therefore, understanding the requirement, we can extend cache-aided NOMA for joint radar sensing and communications. The NOMA-aided joint radar and communication paradigm has been investigated to empower double spectrum sharing, where superimposed multicast and unicast communication signals have been exploited as radar probing waveforms \cite{mu2022noma }. A beamforming design problem of a NOMA-ISAC system has been addressed to maximize the sum throughput for the communication system and enhance the effective sensing power \cite{ Wang2022NOMA }.

\subsubsection{STAR-RIS Network for 360° Coverage}
The only function of reflectors in the conventional IRS systems is to reflect incident signals constructively towards destinations. In this topology, transmitters and receivers need to be on the same side of the reflector, which restricts the flexible employment of the IRS systems. To deal with this shortcoming and facilitate more flexible communication systems, simultaneous transmitting and reflecting RISs (STAR-RISs) can be employed. Unlike conventional IRS reflectors, STAR-RIS divides incident signals into two parts, one part reflects from the surface, and another part propagates into the other side of the RIS. Recently, Mu \textit{et al.} proposed three STAR-RIS operating protocols, namely energy splitting, mode switching, and time switching, and formulated a power consumption minimization problem for all the protocols under the constraint of data rate \cite{mu2021simultaneously}. However, the NOMA and cache-aide NOMA in the STAR-RIS research domain is yet not explored and could be an evolutionary application for next-generation communications.

\subsubsection{NOMA-Empowered Robotic Users} Future human societies in different fields will be surrounded by the application of robotic techniques from smart homes to smart factories. Instead of operating robots on their self-centered individual computational units, robots can be connected with wireless networks as robotic users and operated exchanging information with the APs \cite{liu2021robotic}. The challenges associated with the application areas of robotic users make difficulties in resource management. Operating large numbers of robots is much more challenging, especially when they are deployed for different tasks. To deal with this scenario, researchers have recently started initial research work to investigate the capability of NOMA in robotic communications \cite{mu2021intelligent}.

\subsubsection{Orthogonal Time Frequency Space (OTFS)-NOMA} 
Providing communication maintaining 5G standards to various types of users with different mobility profiles is one of the essential objectives of 5G and B5G systems. Doppler frequency shifts and frequent channel estimation with reliability are two central challenges for high-mobility users. Doppler frequency shift introduces inter-carrier interference, and channel parameters realization timely causes additional system overhead. Orthogonal time-frequency space (OTFS) modulation has been proposed recently to encounter high mobility-related issues \cite{raviteja2018practical}. In OTFS, the primary task is to place high-mobility users' signals in a delay-Doppler plane and converts the channels that are time-varying in the time-frequency plane to time-invariant channels in the delay-Doppler plane. As a result, delay-Doppler plane can directly estimate the channel parameters. Now consider a scenario when highly mobile users occupy the bandwidth resources and time slots, and the users do not require a high data rate or channel gain quality is poor. In this case, the spectral efficiency of OTFS may be low and NOMA-based OTFS can be a solution to this. In \cite{ding2019otfs}, the authors proposed OTFS-NOMA to improve the spectral efficiency and delivery latency with heterogeneous mobility profiles. The OTFS-NOMA technique groups users with different mobility for implementing the NOMA principle. In this domain, NOMA and cache have not been explored much. 

\subsubsection{NOMA-aided Internet of Health}
Internet of health (IoH) is steadily emerging as a necessary service for human health monitoring under the forthcoming 6G communications. IoH services are required to communicate with a massive number patients' electronic medical devices to improve their quality of health. IoH services include real-time remote diagnosis, remote treatment (telemedicine), and remote surgery for emergencies. NOMA is a promising candidate capable of simultaneously transmitting information to multiple patients maintaining coordination among numerous smart devices. To establish in-home medical networks, Xuewan \textit{et al.} proposed a multi-carrier NOMA framework that connects comparatively more monitoring units and transfers more information bits compared to OMA-based designs \cite{Zhang2021Sparse}. Based on patients' medical history, some medical advices as first aid services can be cached in the health monitoring unit. Cache enabled NOMA is particularly useful because, often, the medical information is data heavy (high resolution tomography or video) and local storage of patient data is useful considering their limited mobility.
\subsubsection{NOMA-aided Visible Light Communications} For short-range communications, visible light communication (VLC) is a promising communication scheme operated through an unlicensed spectrum with high secrecy and low energy. Efficient MA techniques are required to be implemented for VLC systems to improve spectral efficiency since the modulation bandwidth is narrow in the VLC. NOMA could be an attractive MA for the VLC. Each transmitter of practical VLC serves has a limited number of users which leads to a reduced SIC-based NOMA decoding process keeping control of the outage probabilities. Furthermore, the slow-varying channel characteristic of the VLC reduces the overhead and complexity required for accurate CSI estimation at the NOMA transmitter. These two unique features of the VLC make the NOMA-based VLC advantageous over the OMA-VLC in terms of the ergodic sum-rate \cite{Yin2016Performance}. MIMO-NOMA-based VLC is also becoming a popular approach to enhance the sum-rate \cite{Chen2018OnThe}. Depending on the requirement and network model, the NOMA-based VLC can be extended to cache-aided NOMA-based VLC systems.
\subsubsection{Hybrid Free Space Optical Communication}
Like the VLC, free-space optical (FSO) communication is another short-range, line-of-sight communication system exploiting the optical domain for the next generation. The FSO communication system fundamentally works on intensity modulation of laser and direct detection by a photodetector in the near-infrared range ($750- 1600$ nm) and transmits data in Gbps range through high bandwidth optical channels. The FSO communication uses a laser as a transmitter with low implementation cost, and the directional communication nature provides higher security compared to traditional communication systems. Despite so many advantages, weather conditions, atmospheric turbulence, and pointing errors (misalignment between the optical transmitter and receiver) are the inescapable challenges of FSO systems. In \cite{ Jamali2020Uplink } the authors have analyzed the performance of a mixed RF-FSO system and validated the superiority of the FSO backhauling and high-reliability NOMA systems over the conventional RF backhauling. Cooperative relay transmission capability can make NOMA the most suitable technology for hybrid FSO/ radio frequency communication systems. Cache-aided NOMA systems are particularly attractive for hybrid RF/FSO systems as cache can help in alleviating issues like difference of data rates over parallel RF and FSO links.

\section{Conclusion}\label{sec: conc}
This article presented a comprehensive survey and reviewed the state-of-the-art research contributions of the cache-aided NOMA technique in wireless communications. First, we explained the fundamental concepts, operating principles, and major challenges associated with the cache and NOMA techniques. We then flexibly amalgamated cache with NOMA and discussed the motivations and goals of cache-aided NOMA-based wireless networks. This article explicitly presented the primary considerations related to cache-aided NOMA network design, including cache memory allocation, most popular content selection, and optimal content placement. This survey paper thoroughly reviewed the frameworks for achieving primary goals such as the increased probability of successful decoding, sum-rate maximization, delay reduction, interference cancellation, and energy consumption minimization. We found efficient placement of popular contents and suitable power allocation are the two essential requirements for designing cache-aided NOMA systems. Next, we categorized the research articles depending on the application scenarios of cache-aided NOMA systems. We then identified that the cache-aided NOMA technology is mostly applied in vehicular communications, UAV-based networks, MEC, and cellular communications. Advantages and challenges related to the utilization of cache-aided NOMA technology in these scenarios are presented. We discussed the benefits of using cache-aided NOMA in these scenarios, and concluded that balancing the performance and energy consumption is the most challenging tasks. Finally, we highlighted existing open challenges and future research directions of the cache-aided NOMA technology.

\bibliographystyle{./IEEEtran}    
\bibliography{IEEEabrv,./cache_survey}

\begin{IEEEbiography}[{\includegraphics[width=1in,height=1.25in,clip,keepaspectratio]{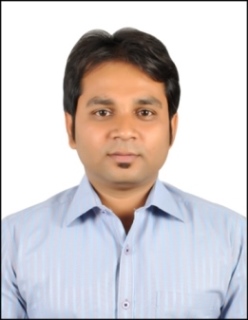}}]{Dipen Bepari} received his B-Tech degree in Electronics and Communication Engineering from Jalpaiguri Government Engineering College, West Bengal, India, and completed M.Tech. from National Institute of Technology, Durgapur, India. He has received Ph.D. degree from Department of Electronics Engineering, IIT (ISM), Dhanbad, India. He received scholarship from the University Grant Commission (UGC), Government of India for the period 2008–2010  during M.Tech., and from the Ministry of Human Resource and Development (MHRD), Government of India for the period 2012–2017 during Ph.D. Presently he is working at National Institute of Technology, Raipur, India. His research interests include cognitive radio networks, wireless sensor networks, energy harvesting, and NOMA technique. 
\end{IEEEbiography}

\begin{IEEEbiography}[{\includegraphics[width=1in,height=1.25in,clip,keepaspectratio]{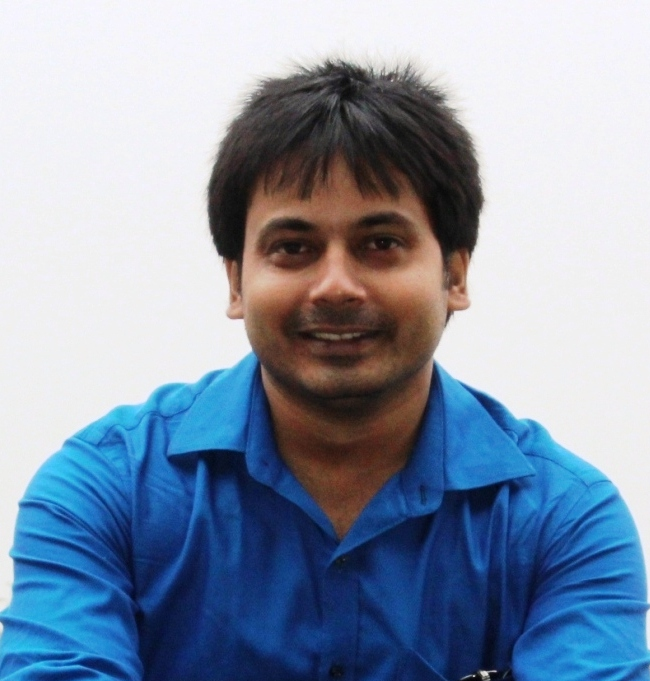}}]{Soumen Mondal} (S'16-S'20) received his B.Tech degree in Electronics and Communication Engineering in 2008 from Haldia Institute of Technology, Haldia, India, M.Tech. degree in Telecommunication Engineering in 2010, and Ph.D. degree in 2021 from National Institute of Technology, Durgapur, India. Dr. Mondal has published about 15 research papers in refereed journals. His research interests include Cognitive Radio Networks, Energy Harvesting, Intelligent Reflecting Surface, FSO and NOMA.
\end{IEEEbiography}

\begin{IEEEbiography}[{\includegraphics[width=1in,height=1.25in,clip,keepaspectratio]{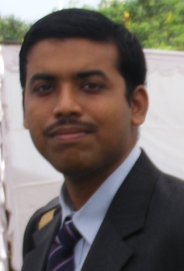}}]{Aniruddha Chandra} (M'08--SM'16) received BE, ME, and PhD degrees from Jadavpur University, Kolkata, India, in 2003, 2005 and 2011, respectively. 

He joined the Electronics and Communication Engineering Department, National Institute of Technology, Durgapur, India, in 2005. He is currently an Associate Professor there. In 2011, he was a Visiting Lecturer at the Asian Institute of Technology, Bangkok. From 2014 to 2016, he worked as a Marie Curie fellow at Brno University of Technology, Czech Republic. In 2019, he worked as a Visiting Researcher at the Slovak University of Technology, Slovakia. In 2022, he was a Guest Researcher at Niigata University, Japan.

Dr. Chandra has published about 130 research papers in refereed journals and peer-reviewed conferences. He is a co-recipient of the best short paper award at IEEE VNC 2014, held in Paderborn, Germany, and delivered a keynote lecture at IEEE MNCApps 2012, held in Bangalore, India. He is currently the secretary of the IEEE P2982 Standard working group and IEEE ComSoc RCC SIG on Propagation Channels for 5G and Beyond. His primary area of research is physical layer issues in wireless communication.
\end{IEEEbiography}

\begin{IEEEbiography}[{\includegraphics[width=1in,height=1.25in,clip,keepaspectratio]{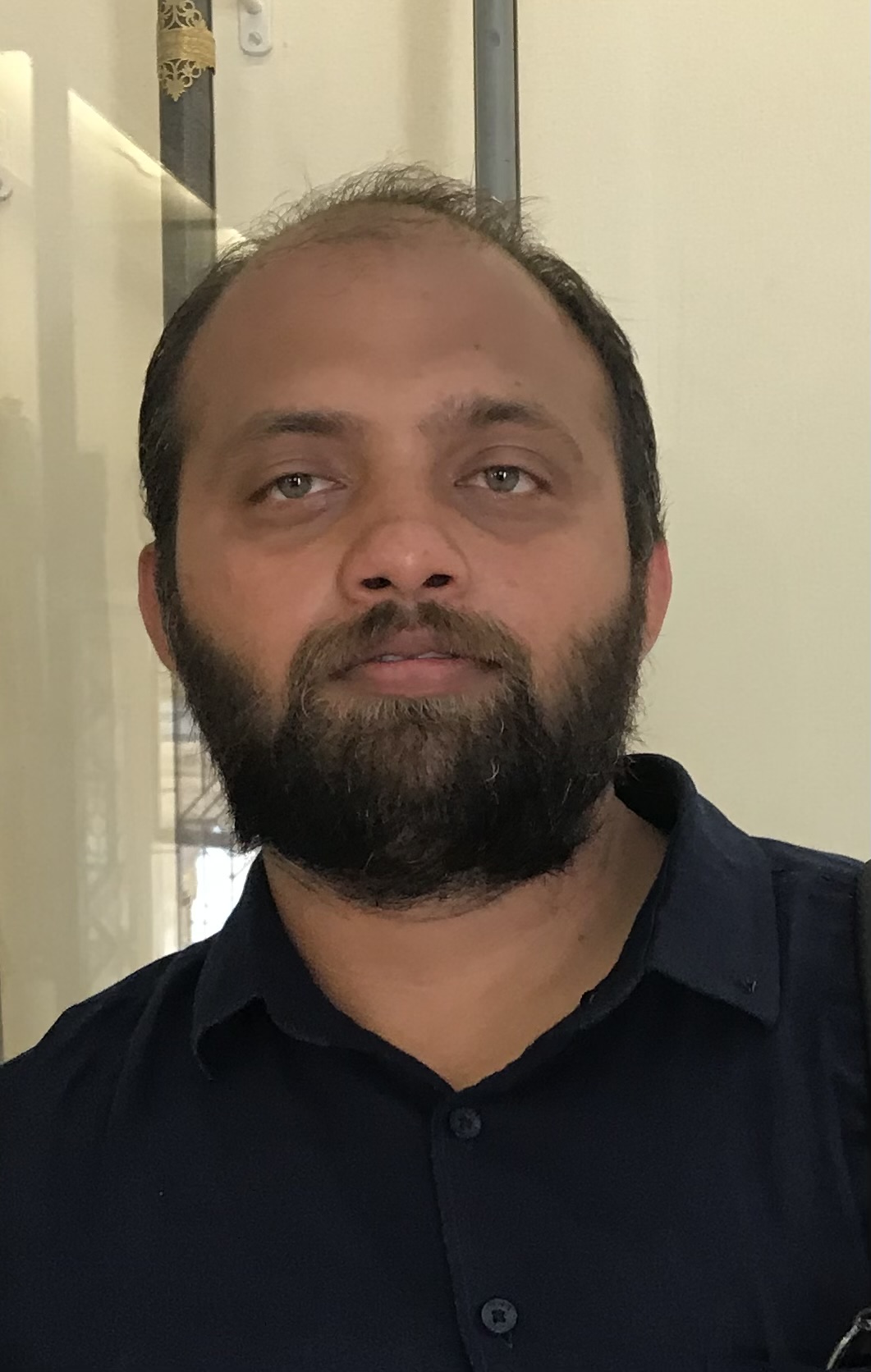}}]{Rajeev Shukla} (S'20) received his BE degree in Electronics and Telecommunication Engineering from Rungta College of Engineering and Technology in 2011 and ME degree in Communication Engineering from Chhattrapati Shivaji Institute of Engineering and Technology in 2015. He is currently pursuing his PhD degree in Wireless Communication from National Institute of Technology, Durgapur, India. He has worked as an Assistant Professor from 2012 to 2015 in School of Engineering, MATS University, Raipur and from 2016 to 2018 in Shekhawati Institute of Engineering and Technology, Sikar. From 2018 to 2020, he worked as a Junior Research Fellow at Indian Institute of Technology, Jodhpur, where he worked on a project on SAR image processing. His research interests include 5G, millimeter waves, channel modelling, and cognitive radio.
\end{IEEEbiography}

\begin{IEEEbiography}[{\includegraphics[width=1.05in,height=1.35in,clip,keepaspectratio]{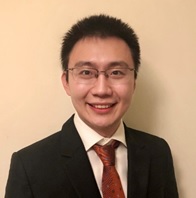}}]{Yuanwei Liu} (S’13–M’16–SM’19, http://www.eecs.qmul.ac.uk/~yuanwei) received the B.S.  and M.S. degrees from the Beijing University of  Posts and Telecommunications in 2011 and 2014,  respectively, and the PhD degree in electrical engineering from the Queen Mary University of London,  U.K., in 2016. He was with the Department of Informatics, King’s  College London, from 2016 to 2017, where he was a Post-Doctoral Research Fellow. He has been a Senior Lecturer (Associate Professor) with the School of Electronic Engineering and Computer Science, Queen Mary University of London, since Aug. 2021, where he was a Lecturer (Assistant Professor) from 2017 to 2021.  His research interests include non-orthogonal multiple access, 5G/6G networks, RIS, integrated sensing and communications,  and machine learning. 

Yuanwei Liu is a Web of Science Highly Cited Researcher 2021. He is currently a Senior Editor of  IEEE Communications Letters,  an Editor of the IEEE Transactions on Wireless Communications and the IEEE Transactions on Communications. He serves as the leading Guest Editor for IEEE JSAC special issue on Next Generation Multiple Access, a Guest Editor for IEEE JSTSP special issue on Signal Processing Advances for Non-Orthogonal Multiple Access in Next Generation Wireless Networks. He received IEEE ComSoc Outstanding Young Researcher Award for EMEA in 2020. He received the 2020 IEEE Signal Processing and Computing for Communications (SPCC) Technical Early Achievement Award, IEEE Communication Theory Technical Committee (CTTC) 2021 Early Achievement Award. He received IEEE ComSoc Young Professional Outstanding Nominee Award in 2021. He has served as the Publicity Co-Chair for VTC 2019-Fall. He is the leading contributor for “Best Readings for Non-Orthogonal Multiple Access (NOMA)” and the primary contributor for “Best Readings for Reconfigurable Intelligent Surfaces (RIS)”. He serves as the chair of Special Interest Group (SIG) in SPCC Technical Committee on the topic of signal processing Techniques for next generation multiple access (NGMA), the vice-chair of SIG Wireless Communications Technical Committee (WTC) on the topic of Reconfigurable Intelligent Surfaces for Smart Radio Environments (RISE), and the Tutorials and Invited Presentations Officer for Reconfigurable Intelligent Surfaces Emerging Technology Initiative.
\end{IEEEbiography}

\begin{IEEEbiography}[{\includegraphics[width=1in,height=1.25in,clip,keepaspectratio]{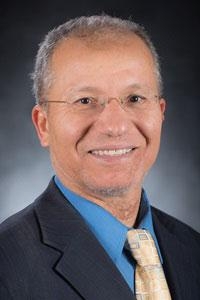}}]{Mohsen Guizani}(M’89–SM’99–F’09) received the BS (with distinction), MS and PhD degrees in Electrical and Computer engineering from Syracuse University, Syracuse, NY, USA in 1985, 1987 and 1990, respectively. He is currently a Professor of Machine Learning and the Associate Provost at Mohamed Bin Zayed University of Artificial Intelligence (MBZUAI), Abu Dhabi, UAE. Previously, he worked in different institutions in the USA. His research interests include applied machine learning and artificial intelligence, Internet of Things (IoT), intelligent autonomous systems, smart city, and cybersecurity. He was elevated to the IEEE Fellow in 2009 and was listed as a Clarivate Analytics Highly Cited Researcher in Computer Science in 2019, 2020 and 2021. Dr. Guizani has won several research awards including the “2015 IEEE Communications Society Best Survey Paper Award”, the Best ComSoc Journal Paper Award in 2021 as well five Best Paper Awards from ICC and Globecom Conferences. He is the author of ten books and more than 800 publications. He is also the recipient of the 2017 IEEE Communications Society Wireless Technical Committee (WTC) Recognition Award, the 2018 AdHoc Technical Committee Recognition Award, and the 2019 IEEE Communications and Information Security Technical Recognition (CISTC) Award. He served as the Editor-in-Chief of IEEE Network and is currently serving on the Editorial Boards of many IEEE Transactions and Magazines. He was the Chair of the IEEE Communications Society Wireless Technical Committee and the Chair of the TAOS Technical Committee. He served as the IEEE Computer Society Distinguished Speaker and is currently the IEEE ComSoc Distinguished Lecturer.
\end{IEEEbiography}
\begin{IEEEbiography}[{\includegraphics[width=1in,height=1.25in,clip,keepaspectratio]{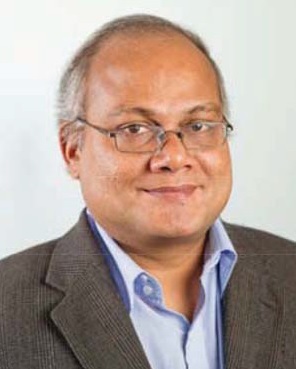}}]{Arumugam Nallanathan} (S'97-–M'00-–SM'05-–F'17)  has been a Professor of Wireless Communications and the Head of the Communication Systems Research Group, School of Electronic Engineering and Computer Science, Queen Mary University of London, since September 2017. He was with the Department of Informatics, King’s College London from December 2007 to August 2017, where he was a Professor of Wireless Communications from April 2013 to August 2017 and a Visiting Professor from
September 2017. He was an Assistant Professor with the Department of Electrical and Computer Engineering, National University of Singapore from August 2000 to December 2007. He published nearly 500 technical papers in scientific journals and international conferences. His research interests include artificial intelligence for wireless systems, B5G wireless networks, Internet of Things, and molecular communications.

He is a co-recipient of the Best Paper Awards presented at the IEEE Communications Society SPCE Outstanding Service Award 2012, the IEEE Communications Society RCC Outstanding Service Award 2014, the IEEE International Conference on Communications in 2016, IEEE Global Communications Conference 2017, and the IEEE Vehicular Technology Conference in 2018. He served as the Chair for the Signal Processing and Communication Electronics Technical Committee of IEEE Communications Society and Technical Program Chair and member of Technical Program Committees in numerous IEEE conferences. He is an Editor-at-Large for IEEE TRANSACTIONS ON COMMUNICATIONS and a Senior Editor for IEEE WIRELESS COMMUNICATIONS LETTERS. He was an Editor for IEEE TRANSACTIONS ON WIRELESS COMMUNICATIONS from 2006 to 2011, IEEE TRANSACTIONS ON VEHICULAR TECHNOLOGY from 2006 to 2017, and IEEE SIGNAL PROCESSING LETTERS. He has been selected as a Web of Science Highly Cited Researcher in 2016 and an AI 2000 Internet of Things Most Influential Scholar in 2020. He is an IEEE Distinguished Lecturer.
\end{IEEEbiography}

\end{document}